\documentclass[english,11pt,nofootinbib,showpacs,notitlepage,usenames,dvipsnames,svgnames,table]{revtex4-2}%{article}
\usepackage[T1]{fontenc}
\usepackage{babel}
\usepackage[utf8]{inputenc}
\usepackage{babel,amsmath,amssymb,mathrsfs,amsfonts,calligra,euscript,textcomp,graphicx,wrapfig,fontenc,pst-blur,tabu,lipsum,%
float,empheq,pst-node,environ,framed,ccaption,capt-of,enumitem,subcaption,enumitem,framed,txfonts,lmodern,cancel,diagbox,multirow,float,tabularx}
\usepackage[section]{placeins}
\definecolor{darkgreen}{rgb}{0,.7,0}
\usepackage[colorlinks,linkcolor=blue,citecolor=green,pagebackref=true]{hyperref}
\def\dac{\displaystyle\frac}

\def\[{\left[}
\def\]{\right]}
\def\({\left(}
\def\){\right)}

\def\1{{\bf CI}}
\def\2{{\bf CII}}
\def\3{{\bf CIII}}

\def\c{\mathcal{C}^1}

\newcommand{\eq}[1]{\begin{equation}#1\end{equation}}

\newcommand{\diag}{\mathop{\rm diag}\nolimits}
\newcommand{\const}{\mathop{\rm const}\nolimits}

\newenvironment{tightcenter}{%
  \setlength\topsep{0pt}
  \setlength\parskip{0pt}
  \begin{center}
}{%
  \end{center}
}
\usepackage{accents}

\usepackage{stackengine}
\stackMath
\newcommand\tsup[2][2]{%
 \def\useanchorwidth{T}%
  \ifnum#1>1%
    \stackon[-.5pt]{\tsup[\numexpr#1-1\relax]{#2}}{\scriptscriptstyle\sim}%
  \else%
    \stackon[.5pt]{#2}{\scriptscriptstyle\sim}%
  \fi%
}

\everymath{\displaystyle}

\begin{document}

\baselineskip7mm

\title{Cosmological solutions in Einstein-Gauss-Bonnet gravity with static curved extra dimensions}

\author{Dmitry Chirkov}
\affiliation{Sternberg Astronomical Institute, Moscow State University, Moscow, Russia}
\affiliation{Bauman Moscow State Technical University, Moscow, Russia}
\author{Alex Giacomini}
\affiliation{Instituto de Ciencias F\'isicas y Matem\'aticas, Universidad Austral de Chile, Valdivia, Chile}
\author{Sergey A. Pavluchenko}
\affiliation{Programa de P\'os-Gradua\noexpand\c c\~ao em F\'isica, Universidade Federal do Maranh\~ao (UFMA), 65085-580, S\~ao Lu\'is, Maranh\~ao, Brazil}
\author{Alexey Toporensky}
\affiliation{Sternberg Astronomical Institute, Moscow State University, Moscow 119991 Russia}
\affiliation{Kazan Federal University, Kremlevskaya 18, Kazan 420008, Russia}

\begin{abstract}
In this paper we perform systematic investigation of all possible solutions with static compact extra dimensions and expanding three-dimensional subspace (``our Universe''). Unlike previous papers,
we consider extra-dimensional subspace to be constant-curvature manifold with both signs of spatial curvature. We provide a scheme how to build solutions in all possible number of extra dimensions and perform stability analysis for the solutions found. Our study suggests that the solutions with negative spatial curvature of extra dimensions are always stable while those with positive curvature
are stable for a narrow range of the parameters and the width of this range shrinks with growth of the number of extra dimensions. This explains why in the previous papers we detected compactification
in the case of negative curvature but the case of positive curvature remained undiscovered. Another interesting feature which distinguish cases with positive and negative curvatures is that the latter
do not coexist with maximally-symmetric solutions (leading to ``geometric frustration'' of a sort) while the former could -- this difference is noted and discussed.
\end{abstract}


%04.20 - General Relativity
%04.20.Dw    Singularities and cosmic censorship
%04.20.Fy    Canonical formalism, Lagrangians, and variational principles
%04.20.Jb    Exact solutions

%04.25.dc    Numerical studies of critical behavior, singularities, and cosmic censorship

%04.50.-h    Higher-dimensional gravity and other theories of gravity
%04.50.Gh    Higher-dimensional black holes, black strings, and related objects
%04.50.Kd    Modified theories of gravity

%11.25Mj     Compactifications and four-dimensional models

%98.80.-k    Cosmology
%98.80.Bp    Origin and formation of the Universe
%98.80.Cq    Particle-theory and field-theory models of the early Universe

\maketitle

\section{Introduction}

Einstein-Gauss-Bonnet Gravity (EGB) is the simplest example of a larger family of gravity theories known as Lovelock Gravities~\cite{LL}. Lovelock gravities are characterized by the fact that their actions possess higher power curvature terms but whose variation lead to equations of motion which remain of second order derivative in the metric.
Lovelock gravities are therefore the most natural generalization of General Relativity to higher space-time dimensions. EGB gravity whose action has additional term quadratic in the curvature with respect to General Relativity exists for space-time dimensions $d\geq 5$ (in four dimensions the quadratic Gauss Bonnet term does not affect the equations of motion being topological). It is the most studied Lovelock gravity in literature as it has many of the features of more generic Lovelock gravities but keeping a relatively simple action. Moreover EGB gravity can also be seen as the low energy limit of certain string theories~\cite{Zwie}.

EGB gravity, whose action holds $\Lambda$-term,  Einstein-Hilbert term and quadratic Gauss-Bonnet term, has the remarkable feature that it can possess up to two independent maximally symmetric solutions (even with different sign of the curvature scale). Indeed, in opposition to General Relativity, the curvature scale of the maximally symmetric solutions is not determined only by the $\Lambda$-term but is a function of all three couplings of the theory. This can be immediately seen by plugging
the {\it ansatz} of a maximally symmetric space-time in the equations of motion of EGB gravity. One gets a quadratic equation in the scale and therefore up to two independent
maximally symmetric solutions. Indeed also for the EGB black hole solutions one gets in general two branches with different asymptotic behavior~\cite{BD}.
However the discriminant of the quadratic equation for the curvature scale can also be imaginary for a range of values of the coupling constants. In this case
there exist no maximally symmetric space-time solution at all. This situation remained almost unexplored in literature due to the fact that it is difficult to give
reasonable asymptotic falloff conditions in such cases. However this situation of non existing maximally symmetric solutions can be of interest in the context of dynamical compactification in cosmology. Indeed one may wonder why a space-time should tend to a compactified manifold with three large dimensions and $D$ compact dimensions instead of tending
to an isotropic space-time which seems more natural. The non-existence of a maximally-symmetric solution would force the space-time to search for a less symmetric configuration and give a very natural explanation for compactification. This has been explored, to the best knowledge of the authors for the first time in~\cite{CGP1, CGP2, CGPT}.
The situation where the EGB theory does not admit maximally symmetric solutions was called in these papers ``geometric frustration''. The term ``geometric frustration'' is normally used in the context of condensed matter physics when a system can not take the configuration of minimal energy due to topological obstructions. In the previously cited papers, for simplicity it was assumed that space-time had the structure of a warped product of a $4$-dimensional  flat FRW space-time  with a $D$ dimensional constant curvature space with an independent scale factor. In the case that  curvature of the compact dimensions is negative and moreover the couplings of the EGB are chosen from the open region of couplings space where geometric frustration occurs it was shown that realistic compactification scenarios exist where the scale factor of the extra dimensions tends to a constant (i.e. stabilizes).
No realistic compactification scenario was found for the case when the extra dimensions have positive curvature or when there is no geometric frustration.
Considering that the cited papers were focused on a generic number $D$ of extra dimensions and especially to the large $D$ limit it is still possible that some cases with positive curvature of the extra dimensions for some particular value of $D$ have just been overseen.

It is interesting to note that the presence of higher-order curvature terms in the Lagrangian is one of the features of string-inspired theories.
Historically, Scherk and Schwarz~\cite{sch-sch} were the
first to demonstrate the presence of the $R^2$ and
$R_{\mu \nu} R^{\mu \nu}$ terms in the Lagrangian of the Virasoro-Shapiro
model~\cite{VSh1, VSh2}. A presence of curvature-squared term of the $R^{\mu \nu \lambda \rho}
R_{\mu \nu \lambda \rho}$ types was demonstrated~\cite{Candelas_etal} fow the low-energy limit
of the $E_8 \times E_8$ heterotic superstring theory~\cite{Gross_etal} to match the kinetic term
for the Yang-Mills field. Later it was demonstrated~\cite{Zwie} that the only
combination of quadratic terms that leads to a ghost-free nontrivial gravitation
interaction is the Gauss-Bonnet (GB) term:

$$
L_{GB} = L_2 = R_{\mu \nu \lambda \rho} R^{\mu \nu \lambda \rho} - 4 R_{\mu \nu} R^{\mu \nu} + R^2.
$$

\noindent This term, initially discovered
by Lanczos~\cite{Lanczos1, Lanczos2} (therefore it is sometimes referred to
as the Lanczos term) is an Euler topological invariant in (3+1)-dimensional
space-time, but not in (4+1) and higher dimensions.
Zumino~\cite{zumino} extended Zwiebach's result on
higher-than-squared curvature terms, supporting the idea that the low-energy limit of the unified
theory should have a Lagrangian density as a sum of contributions of different powers of curvature. In this regard the Einstein-Gauss-Bonnet (EGB) gravity could be seen as a subcase of more general Lovelock
gravity~\cite{LL}, but in the current paper we restrain ourselves with only quadratic corrections and so to the EGB case.

Generally speaking, all extra-dimensional theories have one thing in common---we need to explain where additional dimensions are ``hiding''---since we do not sense them, at least with the current level of experiments. One of
the possible ways to hide extra dimensions and to recover four-dimensional physics, is to build a so-called ``spontaneous compactification'' solution. Exact static solutions with the metric set as a cross product of a
(3+1)-dimensional manifold and a constant curvature ``inner space'',  were found for the first time in~\cite{add_1}, where (3+1)-dimensional manifold being Minkowski (the generalization for
a constant curvature Lorentzian manifold was done in~\cite{Deruelle2}).
In the context of cosmology, it is more useful to consider spontaneous compactification with the four-dimensional part given by a Friedmann-Robertson-Walker (FRW) metric.
In this case it is also natural to consider the size of the extra dimensions being time dependent rather than static. Indeed,
in~\cite{add_4} it was exactly demonstrated that in order to have more realistic model one needs to consider the dynamical evolution of the extra-dimensional scale factor.
In~\cite{Deruelle2}, the equations of motion with time-dependent scale factors were written down for arbitrary Lovelock order in the special case of spatially flat metric (the results were further proven in~\cite{prd09}).
The results of~\cite{Deruelle2} were further analyzed for the special case of 10 space-time dimensions in~\cite{add_10}.
In~\cite{add_8}, the dynamical compactification was studied with the use of Hamiltonian formalism.
More recently, searches for spontaneous  compactifications were performed in~\cite{add13}, where
the dynamical compactification of the (5+1) Einstein-Gauss-Bonnet model was considered; in~\cite{MO04, MO14} with different metric {\it Ans\"atzen} for scale factors
corresponding to (3+1)- and extra-dimensional parts; and in~\cite{CGP1, CGP2, CGPT} (mentioned above), where general (e.g., without any {\it Ans\"atz}) scale factors and curved manifolds were considered. Also, apart from
cosmology, the recent analysis was focused on
properties of black holes in Gauss-Bonnet~\cite{BD, add_rec_1, add_rec_2, addn_1, addn_2} and Lovelock~\cite{add_rec_3, add_rec_4, addn_3, addn_4, addn_4.1} gravities, features of gravitational collapse in these
theories~\cite{addn_5, addn_6, addn_7}, general features of spherical-symmetric solutions~\cite{addn_8}, and many others.

When we are looking for exact cosmological solutions, two main {\it ans\"atzen} are involved -- power-law and exponential. One of first approaches to power-law solutions in EGB gravity
performed in~\cite{Deruelle1, Deruelle2} and more recently they were studied in~\cite{mpla09, prd09, Ivashchuk, prd10, grg10}. One of the first approaches to the
 exponential solutions was done in~\cite{Is86}, the recent works include~\cite{KPT,Iv-16,ErIvKob-16}. We separately described the exponential solutions with variable~\cite{CPT1}
 and constant~\cite{CST2} volume; let us note~\cite{PT} for the discussion of the link between existence of power-law and exponential solutions as well as for the discussion
 about the physical branches of the  solutions. General scheme for finding exponential solutions in arbitrary dimensions and with arbitrary
 Lovelock contributions taken into account described in~\cite{CPT3}. Deeper investigation revealed that not all of the solutions found in~\cite{CPT3} could be called ``stable''~\cite{my15};
 see also~\cite{iv16} for more general approach to the stability of exponential solutions in EGB gravity.

The simplest case -- when the spatial section is the product of spatially-flat three- and extra-dimensional subspaces,
systematic study of all possible regimes in EGB and partially in cubic Einstein-Lovelock gravity was performed in~\cite{my16a, my18a, my18b, my18c, my16b, my17a}. In particular, for vacuum EGB
case it was done in~\cite{my16a} and reanalyzed in~\cite{my18a}. We also added cubic Lovelock term and analyzed the resulting vacuum Einstein-Lovelock cosmology in~\cite{my18b, my18c}.
Similar analysis for EGB model with $\Lambda$-term was performed in~\cite{my16b, my17a} and reanalyzed in~\cite{my18a}. All these studies demonstrate that there are exponential regimes with
expanding three and contracting extra dimensions and they are not suppressed. On the contrary, it is relatively difficult to reach power-law regime naturally.

In the studies described above we made two important assumptions -- both subspaces (three- and extra-dimensional) were considered to be spatially flat and isotropic. But neither of these conditions could
be called ``natural'', so it is interesting to investigate what happens if we left these conditions? In the Friedmann cosmology
spatial curvature plays important role, for example, positive curvature changes the possibility to reach inflationary asymptotic~\cite{infl1, infl2}.
As we previously mentioned, in EGB gravity the influence of the spatial curvature
was studied in~\cite{CGP1, CGP2}, where we described ``geometric frustration'' regime and further investigated it in~\cite{CGPT}.

We addressed effects of both curvature and anisotropy earlier in~\cite{PT2017}. Particularly, we considered initially totally anisotropic (Bianchi-I-type) $(5+1)$- and $(6+1)$-dimensional models and
numerically studied
their evolution. The former of them has only one stable anisotropic exponential solution -- with expanding three and contracting two dimensions -- and it is the only dynamical attractor of the
system. On the contrary, the latter has two possibilities -- expanding three and contracting three
or expanding four and contracting two dimensions, and depending on the initial conditions we could end up in both of the possibilities.
So that if (in)appropriate exponential solution exists and stable, initially anisotropic Universe could end up with ``wrong'' compactification.

We can also note that apart from vacuum and $\Lambda$-term models we considered models with perfect fluid as a source: initially we considered them in~\cite{KPT}, some deeper studies of
(4+1)-dimensional Bianchi-I case was done in~\cite{prd10} and deeper investigation of power-law regimes in pure GB gravity in~\cite{grg10}. Systematic study for all $D$ was started in~\cite{my18d} for
low $D$ and is currently continued for high $D$ cases.

The aim of this paper is actually to check again the compactification, and to do it also for smaller values of $D$ and see if for some particular value there exist  compactification with stabilized extra dimensions. The interest in checking again the positive curvature in extra dimensions is that from a point of view of Kaluza-Klein theory the positive curvature of extra dimensions allows to include non-abelian interactions. In order to do so we check separately the existence of solutions and then their stability due to the fact that if solution exist but is unstable it is useless from a practical point of view. The stability issue of the solutions was not addressed in the previous papers.

The structure of the manuscript is as follows: first we write down equations of motion and then consider several particular cases which differs from each other (for several low $D$ -- number of
extra dimensions -- equations of motion simplify via dropping some of the terms), ending with general case (which has all possible terms). For each of these cases we obtain a solution,
write down perturbed equations and solve them around the found solution. After all cases described, we summarize the results, discuss them and draw conclusions.

\section{Equations of motion}

We start with the standard Einstein-Gauss-Bonnet Lagrangian in the cosmological background (see, e.g.,~\cite{prd09})

\begin{equation}\label{lagr}
{\cal L} = R + \alpha{\cal L}_{GB} - 2\Lambda,
\end{equation}

\noindent where $R$ is the Ricci scalar, $\Lambda$ is $\Lambda$-term (or boundary term) and ${\cal L}_{GB}$,

\begin{equation}
{\cal L}_{GB} = R_{\mu \nu \alpha \beta} R^{\mu \nu \alpha \beta} - 4,
R_{\mu \nu} R^{\mu \nu} + R^2 \label{lagr1}
\end{equation}

\noindent is the Gauss-Bonnet Lagrangian while $\alpha$ is its coupling constant. We consider space-time to be warped product of  Lorenzian and constant curvature manifolds. The former is
$(3+1)$-dimensional while the latter is $D$-dimensional. Then its metric could be written as

\begin{equation}\label{metric}
g_{\mu\nu} = \diag\{ -1, a^2(t), a^2(t), a(t)^2, b(t)^2, b(t)^2 \chi^2(x_4) \ldots, b^2(t) \displaystyle{\prod\limits_{i=4}^{D-4}\chi^2(x_i)}  \},
\end{equation}

\noindent where $\chi(x) = \sin(x)$ for positive and $\chi(x) = \sinh(x)$ for negative curvature of extra dimensions.

Substituting metric into the Lagrangian (\ref{lagr1}), we calculate and vary it with respect to the metric to obtain equations of motion: constraint and dynamical equations for $a(t)$ and $b(t)$,
respectively:

\small
\begin{equation}\label{dyn.eqs}
\begin{array}{l}
D(D-1) \(H_b^2 + \dac{\gamma_D}{b(t)^2}\) + 6DHH_b + 6H^2 + \alpha\( D(D-1)(D-2)(D-3)\(H_b^2 + \dac{\gamma_D}{b(t)^2}\)^2 + 24DH^3H_b + \right. \\  \left.
+ 12D(D-1)H^2(H_b^2 + \dac{\gamma_D}{b(t)^2}) + 24D(D-1)H^2H_b^2 + 12D(D-1)(D-2)H^2(H_b^2 + \dac{\gamma_D}{b(t)^2})  \) = \Lambda, \\
D(D-1)\(H_b^2 + \dac{\gamma_D}{b(t)^2}\) + 4(\dot H + H^2) + 2D(\dot H_b + H_b^2) + 4DHH_b + 2H^2 + \\ + \alpha\( D(D-1)(D-2)(D-3)\(H_b^2 + \dac{\gamma_D}{b(t)^2}\)^2 + 16D(\dot H + H^2)HH_b +
8D(D-1)(\dot H + H^2)\(H_b^2 + \dac{\gamma_D}{b(t)^2}\) + \right. \\ \left. + 8D(\dot H_b + H_b^2)H^2 + 16D(D-1)(\dot H_b + H_b^2)HH_b + 4D(D-1)(D-2)(\dot H_b + H_b^2)\(H_b^2 + \dac{\gamma_D}{b(t)^2}\) + \right. \\ \left. + 4D(D-1)H^2\(H_b^2 + \dac{\gamma_D}{b(t)^2}\)
+8D(D-1)(\dot H + H^2)H_b^2 + 8D(D-1)(D-2)HH_b\(H_b^2 + \dac{\gamma_D}{b(t)^2}\) \) = \Lambda, \\
(D-1)(D-2)\(H_b^2 + \dac{\gamma_D}{b(t)^2}\) + 6(\dot H + H^2) + 2(D-1)(\dot H_b + H_b^2) + 6(D-1)HH_b + 6H^2 + \\ + \alpha\( (D-1)(D-2)(D-3)(D-4)\(H_b^2 + \dac{\gamma_D}{b(t)^2}\)^2 +
24(\dot H + H^2)H^2 + \right. \\ \left. + 12(D-1)(D-2)(\dot H + H^2)\(H_b^2 + \dac{\gamma_D}{b(t)^2}\) + 48(D-1)(\dot H + H^2)HH_b + 24(D-1)(\dot H_b + H_b^2)H^2 + \right. \\ \left. + 4(D-1)(D-2)(D-3)(\dot H_b + H_b^2)\(H_b^2 + \dac{\gamma_D}{b(t)^2}\) + 24(D-1)(D-2)(\dot H_b + H_b^2)HH_b + \right. \\ \left. + 12(D-1)(D-2)(\dot H + H^2)\(H_b^2 + \dac{\gamma_D}{b(t)^2}\) + 24(D-1)H^3H_b + \right. \\ \left. + 12(D-1)(D-2)(D-3)HH_b\(H_b^2 + \dac{\gamma_D}{b(t)^2}\) +
24(D-1)(D-2)H^2H_b^2 \) = \Lambda,
\end{array}
\end{equation}
\normalsize

\noindent where $H \equiv \dot a(t)/a(t)$ is the Hubble parameter associated with ``ordinary'' space, $H_b \equiv \dot b(t)/b(t)$ is the Hubble parameter associated with extra dimensions and $\gamma_D$ is normalized curvature of extra dimensions.

Equation that defines maximally symmetric solutions reads
\eq{(D+3)(D+2)(D+1)D \alpha^2H^4+(D+3)(D+2)\alpha H^2-\xi=0,\label{isotrop}}
\noindent and cosmologically it corresponds to isotropic solution. It has real solutions iff
\eq{\xi\geqslant-\frac{(D+2)(D+3)}{4D(D+1)}.\label{existing_isotrop}}
where $\xi=\alpha \Lambda$.

\section{(3+2)-dimensional case with curvature}

For $D=2$ system (\ref{dyn.eqs}) takes form

\begin{equation}
\begin{array}{l}
2\dac{(\gamma_D + \dot b(t)^2)}{b(t)^2} + 4\dac{\ddot a(t)}{a(t)} + 4\dac{\ddot b(t)}{b(t)} + 8\dac{\dot a(t) \dot b(t)}{a(t) b(t)} + 2\dac{\dot a(t)^2}{a(t)} + \alpha\left(
16\dac{\ddot a(t)(\gamma_D + \dot b(t)^2)}{a(t) b(t)^2} +  \right. \\ \\ + \left. 32\dac{\ddot a(t) \dot a(t) \dot b(t)}{a(t)^2 b(t)} +
16\dac{\ddot b(t)\dot a(t)^2}{a(t)^2 b(t)} +
8\dac{\dot a(t)^2 (\gamma_D + \dot b(t)^2)}{a(t)^2 b(t)^2} +  \right. \\ \\ + \left. 32\dac{\ddot b(t) \dot a(t) \dot b(t)}{a(t) b(t)^2} +
  16\dac{\dot a(t)^2 \dot b(t)^2}{a(t)^2 b(t)^2}   \right) = \Lambda, \\ \\
 6\dac{\ddot a(t)}{a(t)} + 2\dac{\ddot b(t)}{b(t)} + 6\dac{\dot a(t) \dot b(t)}{a(t) b(t)} + 6\dac{\dot a(t)^2}{a(t)} + \alpha\left(
24\dac{\ddot a(t) \dot a(t)^2}{a(t)^3} + \right. \\ \\  + \left.  48\dac{\ddot a(t) \dot a(t) \dot b(t)}{a(t)^2 b(t)} +
24\dac{\ddot b(t)\dot a(t)^2}{a(t)^2 b(t)} +  24\dac{\dot a(t)^3 \dot b(t)}{a(t)^3 b(t)}   \right) = \Lambda,
\end{array} \label{3p2_sys0}
\end{equation}

\noindent complimented with a constraint equation

\small
\begin{equation}
\begin{array}{l}
2\dac{(\gamma_D + \dot b(t)^2)}{b(t)^2} + 12\dac{\dot a(t) \dot b(t)}{a(t) b(t)} + 6\dac{\dot a(t)^2}{a(t)} + \alpha\left( 24\dac{\dot a(t)^2 (\gamma_D + \dot b(t)^2)}{a(t)^2 b(t)^2} +
 48\dac{\dot a(t)^3 \dot b(t)}{a(t)^3 b(t)}  + 48 \dac{\dot a(t)^2 \dot b(t)^2}{a(t)^2 b(t)^2}
\right) = \Lambda. \\ \\
\end{array} \label{3p2_con0}
\end{equation}
\normalsize

We rewrote the system in terms of scale factors, as we are going to look for solutions with ``stabilized extra dimensions'' (so that the size of extra dimensions naturally becomes constant with
respect to time). On the same time we require that three-dimensional subspace to expand with acceleration (after all, we are looking for a solution which could describe observed Universe), then, the
conditions for such solution to exist are $a(t) = \exp (H_0t)$, $b(t) = b_0 \equiv \const$, and the system (\ref{3p2_sys0})--(\ref{3p2_con0}) takes a form:

\begin{equation}
\begin{array}{l}
\dac{2\gamma_D}{b_0^2} + 6H_0^2 + \dac{24\gamma_D \alpha H_0^2}{b_0^2} = \Lambda, \\
12H_0^2 + 24\alpha H_0^4  = \Lambda.
\end{array} \label{3p2_sys1}
\end{equation}

One can see that three dynamical equations shrink to two -- one of the variables becomes static ($b(t) = b_0 \equiv \const$), so there is no ``dynamical'' equation which corresponds to it anymore. One can
also note that we keep $\gamma_D$ arbitrary, unlike analysis for ``geometric frustration'' regime~\cite{CGP1, CGP2, CGPT}. Now choosing new variables $x=1/b_0^2$ and $y=H_0^2$, we can rewrite
 (\ref{3p2_sys1}) as

\begin{equation}
\begin{array}{l}
2z + 6y + 24\alpha zy = \Lambda, \\
12y + 24\alpha y^2 = \Lambda,
\end{array} \label{3p2_sys2}
\end{equation}

\noindent where we absorbed $\gamma_D$ into $x$: $z = \gamma_D x$. Formally we now can get a direct solution: solve second of (\ref{3p2_sys2}) w.r.t. $y$, substitute resulting $y_\pm$ into first of
(\ref{3p2_sys2}) and get $z_\pm$ for each branch. But to keep analysis consistent with the following sections, dedicated to higher-dimensional cases, where the equations will be cubic (for $(3+3)$) or
even quartic (for higher-dimensional cases), we use same approach as we will be using further. Namely, we introduce $\xi = \alpha\Lambda$ and use scaling of the variables with respect to each other:

\begin{equation}
\begin{array}{l}
\Lambda = \xi/\alpha, ~~~ b_0^2 = \zeta\alpha, ~~~ H_0^2 = \theta/\alpha.
\end{array} \label{3p3_redef1}
\end{equation}

With these redefinitions the system (\ref{3p2_sys2}) takes a form

\begin{equation}
\begin{array}{l}
24\gamma_D\theta + 6\theta\zeta - \xi\zeta + 2\gamma_D =0 , \\
12\theta + 24\theta^2 = \xi.
\end{array} \label{3p2_sys3}
\end{equation}

We can immediately solve the second of it w.r.t. $\xi$:

\begin{equation}
\begin{array}{l}
\xi = 12\theta(2\theta+1),
\end{array} \label{3p2_xi0}
\end{equation}

\noindent and substitute it into the first of (\ref{3p2_sys3}) to get $\zeta$:

\begin{equation}
\begin{array}{l}
\zeta = \dac{\gamma_D (12\theta+1)}{3\theta(4\theta+1)}.
\end{array} \label{3p2_zeta0}
\end{equation}

\begin{figure}
\includegraphics[width=1.0\textwidth, angle=0]{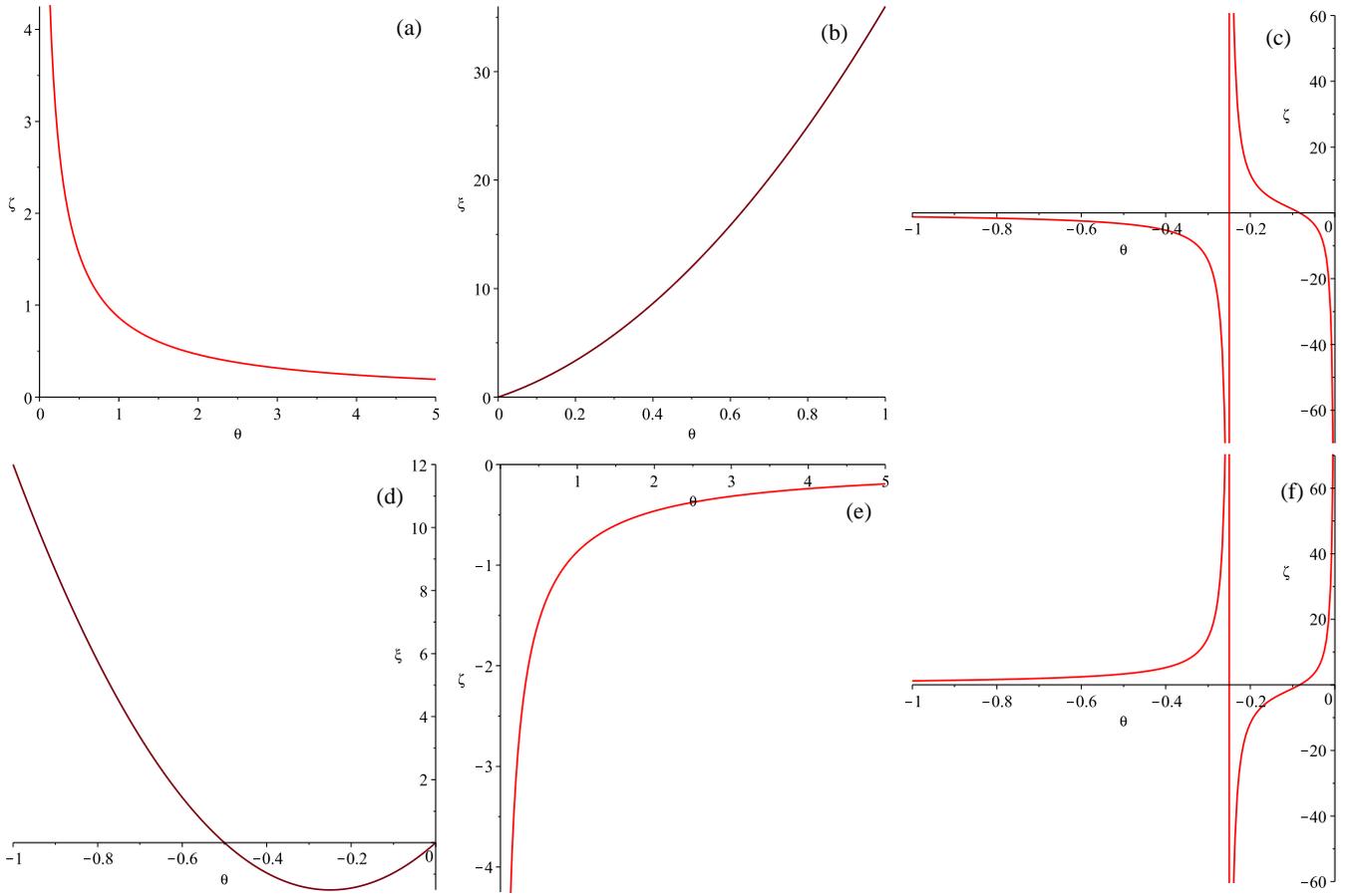}
\caption{Graphs illustrating the behavior of derived functions $\xi$ (\ref{3p2_xi0}) and $\zeta$ (\ref{3p2_zeta0}) for different cases in $(3+2)$-dimensional model: $\alpha > 0$, $\gamma_D > 0$ case: $\zeta(\theta)$ in (a) panel and $\xi(\theta)$ in (b);
$\alpha < 0$, $\gamma_D > 0$ case: $\zeta(\theta)$ in (c) panel and $\xi(\theta)$ in (d); $\zeta(\theta)$ for $\alpha > 0$, $\gamma_D < 0$ case presented in (e) panel while
$\zeta(\theta)$ for $\alpha < 0$, $\gamma_D < 0$ case presented in (f) panel (see the text for more details).}\label{fig_3p2_1}
\end{figure}

Now the initial system is brought to just 1-parametric solution for $\xi$ (\ref{3p2_xi0}) and $\zeta$ (\ref{3p2_zeta0}) with $\theta$ as a parameter. Let us investigate existence of the solutions for
each particular combination of $\{\alpha,\,\gamma_D\}$; the solutions are illustrated in Fig.~\ref{fig_3p2_1}.

{\bf Case 1}: $\alpha > 0$, $\gamma_D > 0$. From definitions (\ref{3p3_redef1}) $\alpha > 0$ means $\theta > 0$ and $\zeta > 0$; from (\ref{3p2_zeta0}) one can see that it is always fulfilled
(see Fig.~\ref{fig_3p2_1}(a)) and from (\ref{3p2_xi0}) one can see that $\xi > 0$ for this case (see Fig.~\ref{fig_3p2_1}(b)).

{\bf Case 2}: $\alpha < 0$, $\gamma_D > 0$. Now $\alpha < 0$ and it means $\theta < 0$ and $\zeta < 0$; from (\ref{3p2_zeta0}) one can see that it is fulfilled for $\theta\in(-\infty; -1/4)\cup(-1/12; 0)$
(see Fig.~\ref{fig_3p2_1}(c)). From (\ref{3p2_xi0}) one can see that $\xi > -3/2$ in this case (see Fig.~\ref{fig_3p2_1}(d)).

{\bf Case 3}: $\alpha > 0$, $\gamma_D < 0$. As in {\bf Case 1}, $\alpha > 0$ means $\theta > 0$ and $\zeta > 0$, but now (\ref{3p2_zeta0}) one can see that it is never fulfilled, which means there ar no
solutions for this case (see Fig.~\ref{fig_3p2_1}(e)).

{\bf Case 4}: $\alpha < 0$, $\gamma_D < 0$. In a way this case is ``opposite'' to the {\bf Case 2}, as now $\theta < 0$ and $\zeta < 0$ are fulfilled for $\theta\in(-1/4; -1/12)$
(see Fig.~\ref{fig_3p2_1}(f)). The corresponding range for $\xi$: $\xi\in(-3/2; -5/6)$ and the graph is the same as in {\bf Case 2} (see Fig.~\ref{fig_3p2_1}(d)).

\subsection{Linear stability of the solutions}

To address the linear stability, we perturb the system (\ref{3p2_sys0}) around the solution with stabilized extra dimensions ($a(t) = \exp (H_0t)$, $b(t) = b_0 \equiv \const$).
For simplicity, we rewrite the system (\ref{3p2_sys0}) back in terms of Hubble parameter $H(t) = \dot a(t)/a(t)$ (additionally, this effectively diminish number of degrees of freedom by one which is
crucial for our task), and we
perturb the system around solution $H(t) = H_0 + \delta H(t)$, $b(t) = b_0 + \delta b(t)$ with $H_0$ and $b_0$ governed by (\ref{3p2_sys1}).
The resulting system of perturbed equations takes a form

\begin{equation}
\begin{array}{l}
4b_0 (1+4\alpha H_0^2) \ddot\delta b(t) + 8b_0 H_0  (1+4\alpha H_0^2) \dot\delta b(t)  + 2b_0 (6H_0^2 - \Lambda) \delta b(t) + ( 16\alpha\gamma_D + 4b_0^2 ) \dot \delta H(t) + \\ +
12H_0 (4\alpha\gamma_D + b_0^2) \delta H(t) = 0, \\
2( 1+12\alpha H_0^2 ) \ddot\delta b(t) + 6H_0( 1+12\alpha H_0^2 ) \dot\delta b(t) + (12H_0^2( 1+12\alpha H_0^2 ) - \Lambda) \delta b(t) + \\ +
6b_0(1+4\alpha H_0^2) \dot \delta H(t) + 24H_0b_0(1+4\alpha H_0^2) \delta H(t) = 0.
\end{array} \label{3p2_pert1}
\end{equation}

To find the solution of the system in the exponential form we transform it into ``normal modes'' with redefinition of the second derivative $\dot \delta b = \delta y$; then the system (\ref{3p2_pert1})
could be replaced with

\begin{equation}
\begin{array}{l}
\begin{pmatrix}
\dot\delta y\\
\dot \delta H \\
\dot \delta b
\end{pmatrix}
= M
\begin{pmatrix}
\delta y\\
\delta H \\
\delta b
\end{pmatrix}
\end{array} \label{3p2_pert2}
\end{equation}

\noindent with $M$ being matrix made of corresponding coefficients. Then the solutions of the (\ref{3p2_pert2}) could be written in the exponential form with the exponents being eigenvalues of $M$. Then,
for solution to be stable, all these exponents should simultaneously have negative real parts. With use of (\ref{3p3_redef1}), (\ref{3p2_xi0}) and (\ref{3p2_zeta0}) these exponents could be rewritten
in terms of only $\theta$ for each choice of $\alpha$ and $\gamma_D$.

\begin{figure}
\includegraphics[width=1.0\textwidth, angle=0]{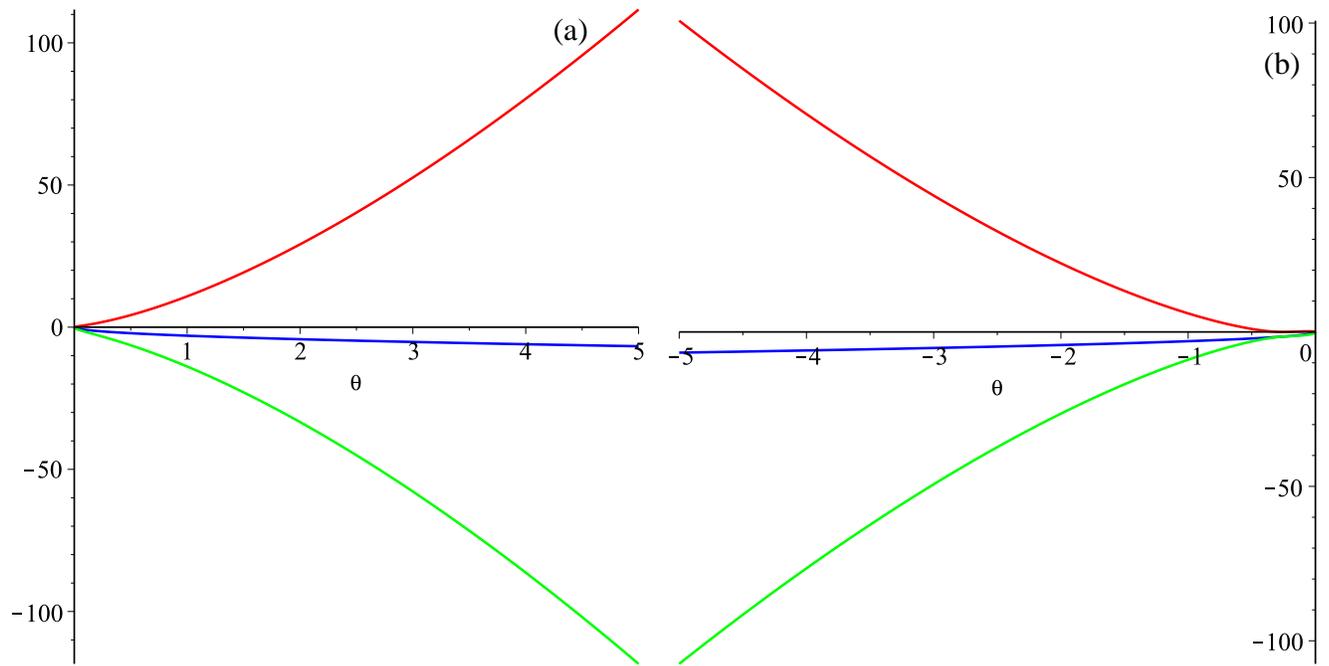}
\caption{Graphs illustrating stability of the solutions for different cases in $(3+2)$-dimensional model: $\alpha > 0$, $\gamma_D > 0$ case in (a) panel;
$\alpha < 0$, $\gamma_D > 0$ and $\alpha < 0$, $\gamma_D < 0$ cases in (b) panel. Different colors correspond to different (three) branches of the eigenvalues
(see the text for more details).}\label{fig_3p2_2}
\end{figure}

Our analysis suggests that for all cases where solutions exist, they are unstable: for {\bf Case 1} ($\alpha > 0$, $\gamma_D > 0$) one of the exponents always has positive real part (see
Fig.~\ref{fig_3p2_2}(a)); same situation is for {\bf Case 2} ($\alpha < 0$, $\gamma_D > 0$) and {\bf Case 4} ($\alpha < 0$, $\gamma_D < 0$), presented in Fig.~\ref{fig_3p2_2}(b). For
{\bf Case 3} ($\alpha > 0$, $\gamma_D < 0$) there are no solutions, as we obtained earlier.

Overall, we can see that in ($3+2$)-dimensional model with curved $2$-dimensional extra-dimensional subspace, there are no stable solutions with stabilizing extra dimensions, regardless of
the curvature of extra-dimensional part.

\section{(3+3)-dimensional case with curvature}

The system of dynamical equations (\ref{dyn.eqs}) for $(3+3)$-dimensional case reads

\begin{equation}
\begin{array}{l}
6\dac{(\gamma_D + \dot b(t)^2)}{b(t)^2} + 4\dac{\ddot a(t)}{a(t)} + 6\dac{\ddot b(t)}{b(t)} + 12\dac{\dot a(t) \dot b(t)}{a(t) b(t)} + 2\dac{\dot a(t)^2}{a(t)} + \alpha\left(
48\dac{\ddot a(t)(\gamma_D + \dot b(t)^2)}{a(t) b(t)^2} +  \right. \\ \\ + \left. 48\dac{\ddot a(t) \dot a(t) \dot b(t)}{a(t)^2 b(t)} +
24\dac{\ddot b(t)\dot a(t)^2}{a(t)^2 b(t)} +
24\dac{\dot a(t)^2 (\gamma_D + \dot b(t)^2)}{a(t)^2 b(t)^2} + 24\dac{\ddot b(t) (\gamma_D + \dot b(t)^2)}{b(t)^3} + \right. \\ \\ + \left. 96\dac{\ddot b(t) \dot a(t) \dot b(t)}{a(t) b(t)^2} +
48\dac{\dot a(t) \dot b(t) (\gamma_D + \dot b(t)^2)}{a(t) b(t)^3} + 48\dac{\dot a(t)^2 \dot b(t)^2}{a(t)^2 b(t)^2}   \right) = \Lambda, \\ \\
2\dac{(\gamma_D + \dot b(t)^2)}{b(t)^2} + 6\dac{\ddot a(t)}{a(t)} + 4\dac{\ddot b(t)}{b(t)} + 12\dac{\dot a(t) \dot b(t)}{a(t) b(t)} + 6\dac{\dot a(t)^2}{a(t)} + \alpha\left(
24\dac{\ddot a(t) \dot a(t)^2}{a(t)^3} + \right. \\ \\ + \left. 24\dac{\ddot a(t)(\gamma_D + \dot b(t)^2)}{a(t) b(t)^2} +   96\dac{\ddot a(t) \dot a(t) \dot b(t)}{a(t)^2 b(t)} +
48\dac{\ddot b(t)\dot a(t)^2}{a(t)^2 b(t)} +
24\dac{\dot a(t)^2 (\gamma_D + \dot b(t)^2)}{a(t)^2 b(t)^2} + \right. \\ \\ + \left. 48\dac{\dot a(t)^3 \dot b(t)}{a(t)^3 b(t)} + 48\dac{\ddot b(t) \dot a(t) \dot b(t)}{a(t) b(t)^2}
 + 48\dac{\dot a(t)^2 \dot b(t)^2}{a(t)^2 b(t)^2}   \right) = \Lambda,
\end{array} \label{3p3_sys0}
\end{equation}

\noindent complimented with a constraint equation

\begin{equation}
\begin{array}{l}
6\dac{(\gamma_D + \dot b(t)^2)}{b(t)^2} + 18\dac{\dot a(t) \dot b(t)}{a(t) b(t)} + 6\dac{\dot a(t)^2}{a(t)} + \alpha\left( 72\dac{\dot a(t)^2 (\gamma_D + \dot b(t)^2)}{a(t)^2 b(t)^2} +
\right. \\ \\ + \left. 72\dac{\dot a(t)^3 \dot b(t)}{a(t)^3 b(t)} + 72\dac{\dot a(t) \dot b(t) (\gamma_D + \dot b(t)^2)}{a(t) b(t)^3} + 144 \dac{\dot a(t)^2 \dot b(t)^2}{a(t)^2 b(t)^2}
\right) = \Lambda. \\ \\
\end{array} \label{3p3_con0}
\end{equation}

The overall procedure is quite similar to the previously described $(3+2)$-dimensional case.
The original system (\ref{3p3_sys0})--(\ref{3p3_con0}) could be brought to the following form under ``stable compactification'' requirement ($a(t) = \exp (H_0t)$, $b(t) = b_0 \equiv \const$):

\begin{equation}
\begin{array}{l}
\dac{6\gamma_D}{b_0^2} + 6H_0^2 + \dac{72\gamma_D \alpha H_0^2}{b_0^2} = \Lambda, \\
\dac{2\gamma_D}{b_0^2} + 12H_0^2 + 24\alpha H_0^4 + \dac{48\gamma_D \alpha H_0^2}{b_0^2} = \Lambda;
\end{array} \label{3p3_sys1}
\end{equation}

\noindent where we kept $\gamma_D$ arbitrary. Choosing new variables $x=1/b_0^2$ and $y=H_0^2$, expressing one of new variables from first of (\ref{3p3_sys1}) and substituting it into the second of
(\ref{3p3_sys1}), we can arrive to a pair of cubic equations:

\begin{equation}
\begin{array}{l}
432\alpha^2 y^3 + 180\alpha y^2 + (15-6\xi)y = \Lambda, \\
864\alpha^2 z^3 + 72\alpha (2\xi+5) z^2 + 30z = \Lambda (2\xi + 3);
\end{array} \label{3p3_sys2}
\end{equation}

\noindent where we absorbed $\gamma_D$ into $x$: $z = \gamma_D x$ and used standard notation $\xi = \alpha\Lambda$. From the definitions used, we should have solutions within $x>0$, $y>0$, and the
analysis in these coordinates becomes quite cumbersome (though not impossible). To simplify things we are going to use different approach -- same as for $(3+2)$-dimensional case. Namely, we use the same
redefinitions (\ref{3p3_redef1}); substituting them into (\ref{3p3_sys1}), the system takes a form

\begin{equation}
\begin{array}{l}
72\gamma_D\theta + 6\theta\zeta - \xi\zeta + 6\gamma_D = 0; \\
24\theta^2\zeta + 48\gamma_D\theta + 12\theta\zeta - \zeta\xi + 2\gamma_D = 0.
\end{array} \label{3p3_sys3}
\end{equation}

And this system has 1-parametric family of solutions:

\begin{equation}
\begin{array}{l}
\xi = \dac{3\theta (144\theta^2 + 60\theta + 5)}{6\theta + 1}, ~~\zeta = \dac{2}{3} \dac{\gamma_D (6\theta+1)}{\theta (4\theta+1)}.
\end{array} \label{3p3_sol1}
\end{equation}

So that given $\alpha$ and $\gamma_D$ we can build all possible solutions for all possible $\theta$. Let us analyze possible solutions and areas of their definitions. To start with, let us notice that
from (\ref{3p3_redef1}) it is clear that for $\alpha > 0$ we should have $\zeta > 0$ and $\theta > 0$ (as both $b_0^2$ and $H_0^2$ cannot be negative) while for $\alpha < 0$ we should have $\zeta < 0$
and $\theta < 0$. Let us consider all four possible combinations of signs for $\alpha$ and $\gamma_D$ separately.

\begin{figure}
\includegraphics[width=1.0\textwidth, angle=0]{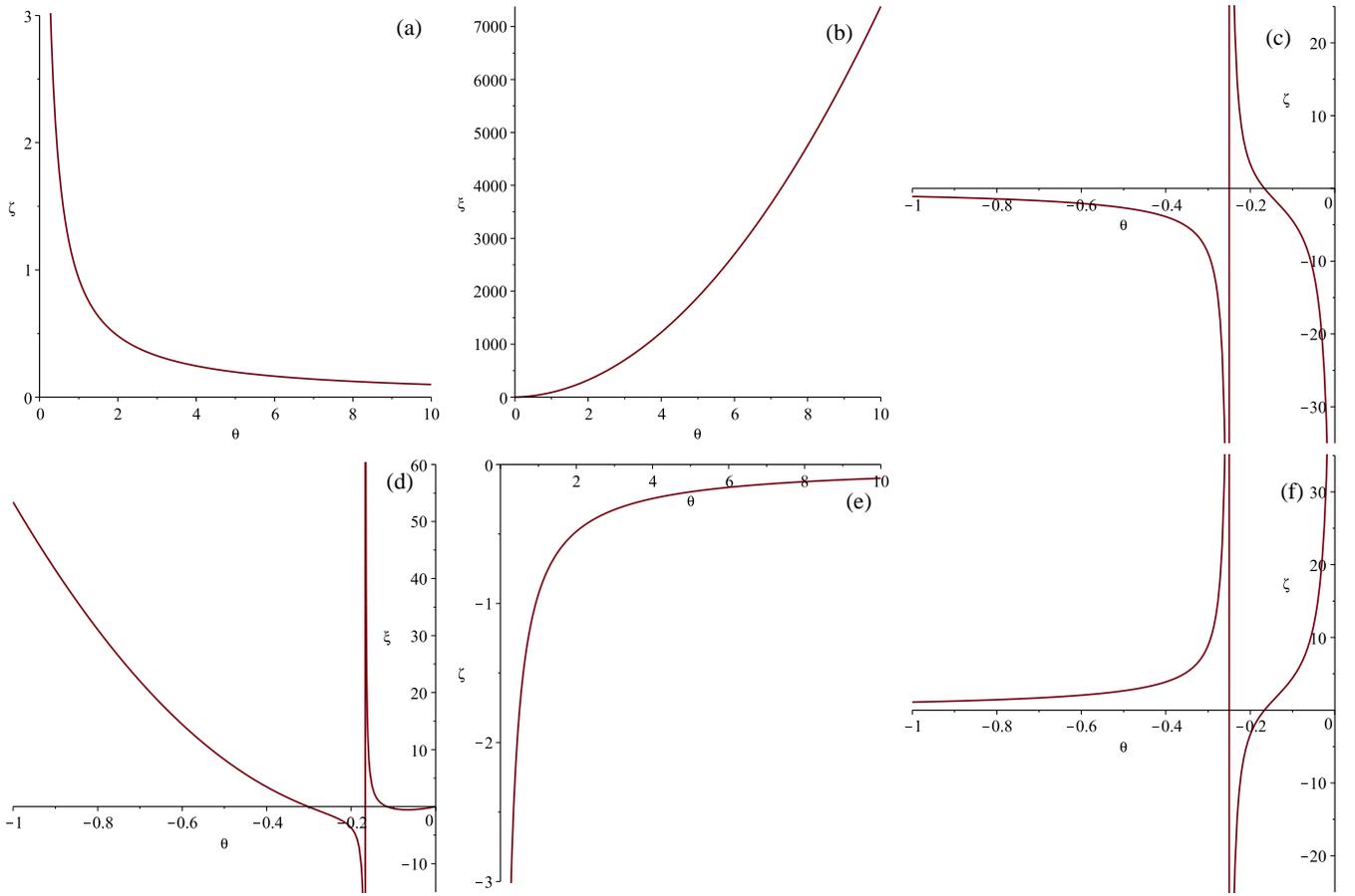}
\caption{Graphs illustrating the behavior of derived functions (\ref{3p3_sol1}) for different cases in $(3+3)$-dimensional model: $\alpha > 0$, $\gamma_D > 0$ case: $\zeta(\theta)$ in (a) panel
and $\xi(\theta)$ in (b);
$\alpha < 0$, $\gamma_D > 0$ case: $\zeta(\theta)$ in (c) panel and $\xi(\theta)$ in (d); $\zeta(\theta)$ for $\alpha > 0$, $\gamma_D < 0$ case presented in (e) panel while
$\zeta(\theta)$ for $\alpha < 0$, $\gamma_D < 0$ case presented in (f) panel (see the text for more details).}\label{fig_3p3_1}
\end{figure}

{\bf Case 1}: $\alpha > 0$, $\gamma_D > 0$. In that case $\theta > 0$ and substituting it into $\zeta$ we can see that $\zeta > 0$ for all $\theta > 0$ (see Fig.~\ref{fig_3p3_1}(a)).
Substituting $\theta > 0$ into $\xi$ we can see that $\xi > 0$ (see Fig.~\ref{fig_3p3_1}(b)). So that for $\alpha > 0$ and $\xi > 0$ there always exist solution with $\gamma_D > 0$.

{\bf Case 2}: $\alpha < 0$, $\gamma_D > 0$. In this case, since $\alpha < 0$, we should consider only $\theta < 0$ and the solution should have $\zeta < 0$, which is satisfied in two regions
(see Fig.~\ref{fig_3p3_1}(c)): $\theta < -1/4$ and $-1/6 < \theta < 0$. Comparing with Fig.~\ref{fig_3p3_1}(d) and performing some basic math, we can learn that within the first range ($\theta < -1/4$)
we have $\xi > 0$ at $\theta < \theta_1$, and $\xi$ is negative at $\theta \in (\theta_1, -1/4)$, with $\xi(-1/4) = -3/2$. Within the second range ($-1/6 < \theta < 0$),
$\xi > 0$ for $\theta \in (-1/6, \theta_2)$ and is negative for
$\theta \in (\theta_2, 0)$. Within the second range we can spot minimum value for $\xi$ which happening at
$\theta_{min} = \sqrt[3]{109 + 6\sqrt{330}}/72 + 1/(72\sqrt[3]{109 + 6\sqrt{330}}) - 11/72 \approx -0.0669$; $\xi(\theta_{min}) \approx -0.5467$. All quoted specific values for $\theta$ are
roots or singularities coming from (\ref{3p3_sol1}); $\theta_1$ and $\theta_2$ ($\theta_1 < \theta_2$) are roots of $(144\theta^2 + 60\theta + 5)=0$:
$\theta_{1,\, 2} = (-5\pm\sqrt{5})/24 \approx \{-0.302,\,-0.115\}$.

To conclude, for $\alpha < 0$ we can have $\gamma_D > 0$
solutions for both $\xi < 0$ and $\xi > 0$.

{\bf Case 3}: $\alpha > 0$, $\gamma_D < 0$. In this case we should have $\theta > 0$ and after plotting $\zeta (\theta)$ in Fig.~\ref{fig_3p3_1}(e), we see that $\zeta < 0$ everywhere which means
that there are no solutions of this kind.

{\bf Case 4}: $\alpha < 0$, $\gamma_D < 0$. In this case we have $\theta < 0$, after plotting $\zeta(\theta)$ in Fig.~\ref{fig_3p3_1}(f) we see that it is ``opposite'' to $\alpha < 0$, $\gamma_D > 0$
case, again we should have $\zeta < 0$ and the area of existence is complimentary to $\alpha < 0$, $\gamma_D > 0$ case -- now it is $\theta \in (-1/4, -1/6)$. Since $\xi(\theta)$ does not depend on
$\gamma_D$, we can use Fig.~\ref{fig_3p3_1}(d) again to find that $\xi < -3/2$ in this case. So that $\gamma_D < 0$ solution could exist only for $\alpha < 0$ and they could be found within
$\theta\in (-1/4, -1/6)$ range.

Now let us analyze linear stability of the obtained solutions.

\subsection{Linear stability of the solutions}

Similarly to the previous case,
to study the linear stability, we perturb the system (\ref{3p3_sys0}) around the solution with stabilized extra dimensions. Since the curvature of the internal subspace is zeroth, we can
rewrite the system (\ref{3p3_sys0}) in terms of Hubble parameter $H(t) = \dot a(t)/a(t)$ (effectively this diminish number of degrees of freedom by one which is crucial for our task), then we
perturb the system around solution $H(t) = H_0 + \delta H(t)$, $b(t) = b_0 + \delta b(t)$ with $H_0$ and $b_0$ governed by (\ref{3p3_sys1}) and pervious section describe a way how to find them.
The resulting system of perturbed equations takes a form

\begin{equation}
\begin{array}{l}
(24\alpha H_0^2 b_0^2 + 24\alpha\gamma_D + 6b_0^2) \ddot\delta b(t) + (48\alpha H_0^3 b_0^2 + 48\alpha H_0\gamma_D + 12 H_0 b_0^2 ) \dot\delta b(t) + \\ + ( 72 \alpha H_0^2 \gamma_D +
18 H_0^2 b_0^2 - 3\Lambda b_0^2 + 6\gamma_D ) \delta b(t) + ( 48\alpha b_0\gamma_D + 4 b_0^3 ) \dot \delta H(t) + \\ + ( 144\alpha H_0 b_0 \gamma_D + 12H_0 b_0^3 ) \delta H(t) = 0, \\
( 48\alpha H_0^2 b_0 + 4b_0 ) \ddot\delta b(t) + ( 144\alpha H_0^3 b_0 + 12H_0 b_0 ) \dot\delta b(t) + (48\alpha H_0^4 b_0 + 24H_0^2 b_0 - 2\Lambda b_0) \delta b(t) + \\ +
(24\alpha H_0^2 b_0^2 + 24\alpha\gamma_D + 6 b_0^2) \dot \delta H(t) + (96\alpha H_0^3 b_0^2 + 96\alpha H_0\gamma_D + 24H_0 b_0^2) \delta H(t) = 0.
\end{array} \label{3p3_pert1}
\end{equation}

\begin{figure}
\includegraphics[width=1.0\textwidth, angle=0]{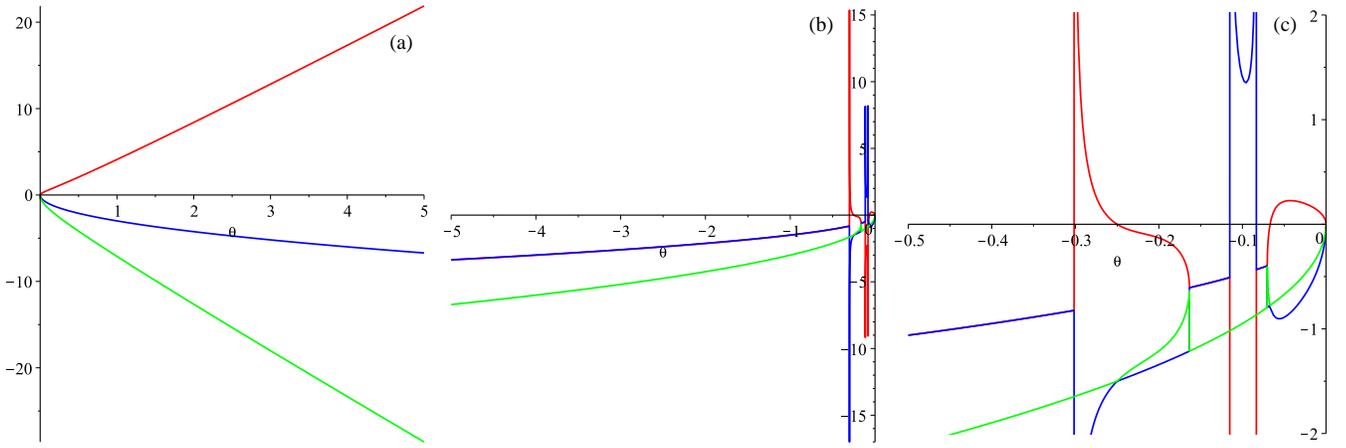}
\caption{Graphs illustrating stability of the solutions for different cases in $(3+3)$-dimensional model: $\alpha > 0$, $\gamma_D > 0$ case in (a) panel;
$\alpha < 0$, $\gamma_D > 0$ and $\alpha < 0$, $\gamma_D < 0$ cases in (b) (``large-scale'') and (c) (``fine structure'') panels.  Different colors correspond to different (three) branches of the
eigenvalues (see the text for more details).}\label{fig_3p3_2}
\end{figure}

 We again use normal modes, and with use of (\ref{3p3_redef1}) and (\ref{3p3_sol1}) the exponents from the eigenvalues could be rewritten in terms of only $\theta$ for each choice of $\alpha$ and $\gamma_D$. So that we examine the exponents for all cases and present our analysis in Fig.~\ref{fig_3p3_2}.

{\bf Case 1}: $\alpha > 0$, $\gamma_D > 0$. Two out of three exponents are real and negative while the third is real and positive for all $\theta$, making this solution unstable
(see Fig.~\ref{fig_3p3_2}(a)). We also remind the reader that for {\bf Case 3} there are no solutions.

{\bf Cases 2 and 4}: $\alpha < 0$  for $\gamma_D > 0$ and $\gamma_D < 0$ respectively -- have the same structure of stability areas so we report them together.
First branch is stable for
$\theta \in (-\infty, \theta_1) \cup (-1/4, \theta_4)$, second branch has negative real part of the exponent everywhere in $\theta < 0$ and the third within
$\theta \in (-\infty, \theta_2) \cup (\theta_5, 0)$. Here $\theta_1$ and $\theta_2$ are the same as defined above while $\theta_3 < \theta_4$ are roots of $(55296\theta^4 + 44352\theta^3 + 12768\theta^2 - 1480\theta + 55) = 0$ -- radicand from the expression and $\theta_5 = -1/12$.
Since we call solution as stable if all three exponents are negative, the overall stability range is the overlap of all three branches; the resulting range is
$\theta\in(-\infty, \theta_1) \cup (-1/4, \theta_2) \cup (\theta_5, \theta_4)$.
 Comparing these ranges with those of existence, we can see that {\bf Case 4} solutions are stable on the entire area of definition ($\theta\in(-1/4, -1/6)$).  On the other hand, {\bf Case 2} solutions
  overlap with stability range for $\theta\in(-\infty, \theta_1) \cup (-1/6, \theta_2) \cup (\theta_5, \theta_4)$. The situation is illustrated in Figs.~\ref{fig_3p3_2}(b--c), where on (b) panel we
  presented ``large-scale'' picture while on (c) panel we focus on the range near zero, where there is ``fine-structure'' of the solutions.

Let us make a note on $\xi$, as it defines $\Lambda$: for {\bf Case 2} both ranges $\theta\in (-\infty, \theta_1)$ and $\theta\in (-1/6, \theta_2)$ give $\xi\in (0, +\infty)$ which, combined with
$\alpha < 0$, gives us $\Lambda < 0$ -- so that these {\bf Case 2} solutions exist for $\alpha < 0$, $\Lambda < 0$. The last range $\theta\in (\theta_5, \theta_4)$ gives $\xi\in (-0.5448, -0.5)$ -- tiny
but negative range, resulting in $\Lambda > 0$.

For {\bf Case 4} we have $\theta\in (-1/4, -1/6)$ which results in $\xi\in (-\infty, -3/2)$, so that $\Lambda > 0$.

\section{(3+4)-dimensional solution}

For $(3+4)$-dimensional case the resulting system for stabilized extra dimensions would be 4th order polynomial, making it even harder to solve explicitly then $(3+3)$-dimensional case. So that
we use same technic as for $(3+3)$ dimensions, namely, we use (\ref{3p3_redef1}) for the resulting system and obtain (after neglecting denominator):

\begin{equation}
\begin{array}{l}
144\gamma_D \theta\zeta + 6\theta\zeta^2 - \xi\zeta^2 + 24\gamma_D^2 + 12\gamma_D \zeta = 0, \\
24\theta^2 \zeta + 144\gamma_D \theta + 12\theta \zeta - \xi\zeta + 6\gamma_D = 0.
\end{array} \label{3p4_sys1}
\end{equation}

We solve the second of (\ref{3p4_sys1}) with respect to $\xi$ and substitute it into the first of (\ref{3p4_sys1}); since we use normalization $\gamma_D = \pm 1$, obviously $\gamma_D^2 = 1$, so the
resulting equation takes a form

\begin{equation}
\begin{array}{l}
24\zeta^2\theta^2 + 6\zeta^2\theta - 6\zeta\gamma_D - 24 = 0,
\end{array} \label{3p4_eq_zeta}
\end{equation}

\noindent which has a solution

\begin{equation}
\begin{array}{l}
\zeta_{\pm} = \dac{3\gamma_D \pm 3 |8\theta + 1|}{6\theta (4\theta + 1)}.
\end{array} \label{3p4_sol_zeta}
\end{equation}

Now we can substitute it into previously found expression for $\xi$ from the second of (\ref{3p4_sys1}):

\begin{equation}
\begin{array}{l}
\xi_{\pm} = \dac{12\theta}{|8\theta+1|\pm\gamma_D} \( \pm 96\gamma_D \theta^2 + 2\theta|8\theta+1| \pm 30\gamma_D \theta + |8\theta+1| \pm 2\gamma_D \).
\end{array} \label{3p4_sol_xi}
\end{equation}

Equations (\ref{3p4_sol_zeta}) and (\ref{3p4_sol_xi}), together with definitions (\ref{3p3_redef1}), completely determine two branches of 1-parametric solution with stabilized extra dimensions in
$(3+4)$-dimensional model. Let us analyze when these solutions exist. As in $(3+3)$-dimensional case, we consider all possible combinations of $\alpha$ and $\gamma_D$ separately.

\begin{figure}
\includegraphics[width=0.9\textwidth, angle=0]{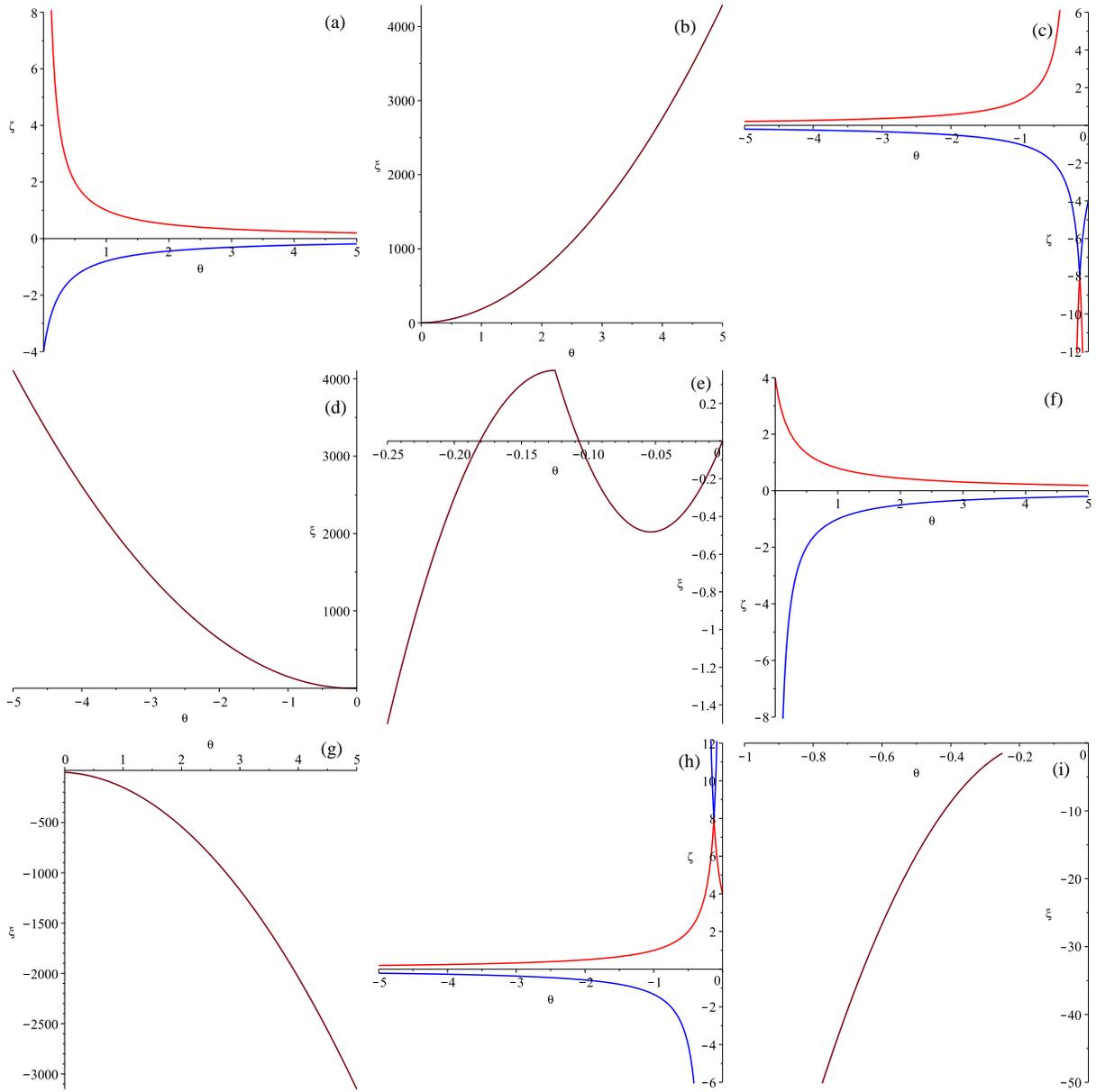}
\caption{Graphs illustrating the behavior of derived functions (\ref{3p3_sol1}) for different cases in $(3+4)$-dimensional model:
$\alpha > 0$, $\gamma_D > 0$ case: $\zeta_\pm(\theta)$ in (a) panel ($\zeta_+$ in red and $\zeta_-$ in blue) and $\xi_+(\theta)$ in (b);
$\alpha < 0$, $\gamma_D > 0$ case: $\zeta_\pm(\theta)$ in (c) panel ($\zeta_+$ in red and $\zeta_-$ in blue), $\xi_-(\theta)$ in (d) and  $\xi_+(\theta)$ in (e);
$\alpha > 0$, $\gamma_D < 0$ case: $\zeta_\pm(\theta)$ in (f) panel ($\zeta_+$ in red and $\zeta_-$ in blue) and $\xi_+(\theta)$ in (g);
$\alpha < 0$, $\gamma_D < 0$ case: $\zeta_\pm(\theta)$ in (h) panel ($\zeta_+$ in red and $\zeta_-$ in blue) and $\xi_-(\theta)$ in (i); (see the text for more details).}\label{fig_3p4_1}
\end{figure}

{\bf Case 1}: $\alpha > 0$, $\gamma_D > 0$. In that case $\theta > 0$ and substituting it into $\zeta_\pm$ we can see that $\zeta_+ > 0$ everywhere in $\theta > 0$ (colored in red in
Fig.~\ref{fig_3p4_1}(a)) while $\zeta_- < 0$ (colored in blue in Fig.~\ref{fig_3p4_1}(a)). So we can conclude that only $\zeta_+$ is viable and plot corresponding $\xi_+$ in Fig.~\ref{fig_3p4_1}(b); from
it one can see that $\xi_+ > 0$ which means $\Lambda > 0$ (since $\alpha > 0$).

{\bf Case 2}: $\alpha < 0$, $\gamma_D > 0$. In that case $\theta < 0$ and substituting it into $\zeta_\pm$ we can see that $\zeta_- < 0$ everywhere in $\theta < 0$ (colored in blue in
Fig.~\ref{fig_3p4_1}(c)) while $\zeta_+ < 0$ only for $\theta > -1/4$ (colored in red in Fig.~\ref{fig_3p4_1}(c)). For ``--'' branch we plot $\xi_-$ in Fig.~\ref{fig_3p4_1}(d) and one can see that
it is always positive (and so that $\Lambda < 0$) while for ``+'' branch $\xi_+$ is plotted in Fig.~\ref{fig_3p4_1}(e) and we can see that $\xi$ could be both positive and negative there. Minimal possible
value is $\xi_+(-1/4) = -3/2$; $\xi_+(\theta)$ hits zero at $\theta_1 = -(5+\sqrt{5})/40 \approx -0.1809$ and $\theta_2 = -3/28 \approx -0.1071$ (and at $\theta = 0$); it has maximum at $\theta=-1/8$
with $\xi_+(-1/8) = 3/8$ and finally it has local minimum at $\theta_{min} = -3/56 \approx -0.05357$ with $\xi_+(\theta_{min}) = -27/56 \approx -0.48214$. So that this case has a variety of parameters
where solutions could possibly exist.

{\bf Case 3}: $\alpha > 0$, $\gamma_D < 0$. In that case $\theta > 0$ and substituting it into $\zeta_\pm$ we can see that $\zeta_+ > 0$ everywhere in $\theta > 0$ (colored in red in
Fig.~\ref{fig_3p4_1}(f)) while $\zeta_- < 0$ (colored in blue in Fig.~\ref{fig_3p4_1}(f)). So we can conclude that only $\zeta_+$ is viable and plot corresponding $\xi_+$ in Fig.~\ref{fig_3p4_1}(g); from
it one can see that $\xi_+ < 0$ which means $\Lambda < 0$ (since $\alpha > 0$). One cannot miss familiarity between Cases 1 and 3 -- ``mirror-like'' behavior of $\zeta_\pm$ branches
(compare Figs.~\ref{fig_3p4_1}(a) and (f)) but since they have different sign for $\gamma_D$, the resulting $\xi$ also have different sign (compare Figs.~\ref{fig_3p4_1}(b) and (g)).

{\bf Case 4}: $\alpha < 0$, $\gamma_D < 0$. In that case $\theta < 0$ and substituting it into $\zeta_\pm$ we can see that $\zeta_+ > 0$ everywhere in $\theta < 0$ (colored in red in
Fig.~\ref{fig_3p4_1}(h)) while $\zeta_- < 0$ only for $\theta < -1/4$ (colored in blue in Fig.~\ref{fig_3p4_1}(h)). So that the only viable branch is $\zeta_-$ and only for $\theta < -1/4$.
The resulting $\xi_-(\theta)$ graph is presented in Fig.~\ref{fig_3p4_1}(i) -- one can clearly see that $\xi < -3/2$ and so $\Lambda > -3/(2\alpha)$.
Again, one cannot miss familiarity between Cases 2 and 4 -- their $\zeta_\pm$ curves are ``mirror-symmetric'' with respect to $\zeta=0$ (compare Figs.~\ref{fig_3p4_1}(c) and (h)) but unlike
Cases 1 and 3 described above, Cases 2 and 4 have more complicated structure which results in quite different branch/range combination for each of them. Also, since now we have only one branch,
for $\xi$ we also have only single choice.

Overall, one can see much more abundant {\it potential} dynamics then for $(3+3)$-dimensional case; let us find out if these solutions are stable or not.

\subsection{Linear stability of (3+4)-dimensional solutions}

\begin{figure}
\includegraphics[width=1.0\textwidth, angle=0]{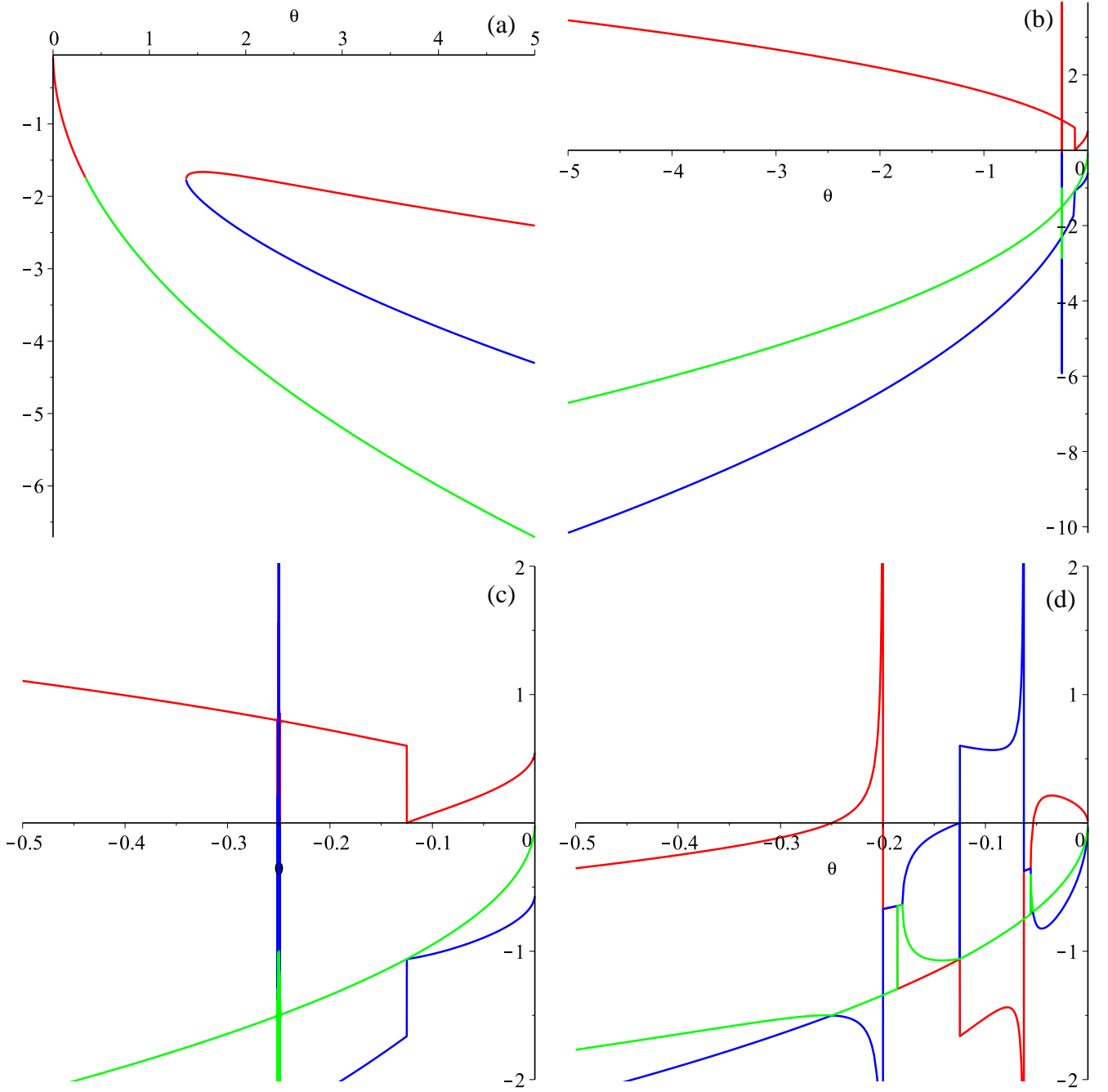}
\caption{Graphs illustrating stability of the solutions for different cases in $(3+4)$-dimensional model: $\alpha > 0$, $\gamma_D < 0$ case in (a) panel;
$\alpha < 0$, $\gamma_D > 0$ with $\zeta_-$ branch in (b) (``large-scale'') and (c) (``fine structure'') panels;
$\alpha < 0$, $\gamma_D > 0$ with $\zeta_+$ branch as well as $\alpha < 0$, $\gamma_D < 0$ with $\zeta_-$ branch in (d) panel (``fine structure'').
Different colors correspond to different (three) branches of the eigenvalues
(see the text for more details).}\label{fig_3p4_2}
\end{figure}

The general scheme for finding stable solutions is exactly the same as in $(3+3)$-dimensional case -- we perturb the system of dynamical equations around the solution, separate perturbations, find
normal modes and investigate when they all simultaneously have negative real parts. Skipping technical details, we report the results for all cases separately; the results are presented in
Fig.~\ref{fig_3p4_2}.

{\bf Case 1}: in this case one of the modes is always real and positive while two others are always real and negative -- overall it means that the solutions are unstable for this case. The situation
resemble same {\bf Case 1} from $(3+3)$-dimensional case so Fig.~\ref{fig_3p3_2}(a) would serve as a good illustration.

{\bf Case 2}: as previously described, here we have possible solutions on both $\zeta_\pm$ branches. For $\zeta_-$ branch one of the modes always has positive real part which makes this branch unstable.
Situation is illustrated in Figs.~\ref{fig_3p4_2}(b--c), with (b) panel presents ``large-scale`` picture and (c) panel -- ``fine-tuning'' at small values of $\theta$ where the structure is complicated.
On the contrary, $\zeta_+$ has all three  modes with negative real parts at $\theta\in (-\infty, -1/4) \cup (-1/5, -1/8) \cup (-1/16, -3/56)$ (see Fig.~\ref{fig_3p4_2}(d)).
If we unite it with the region of existence described
earlier, we will have to drop the first range, resulting in $\theta\in(-1/5, -1/8) \cup (-1/16, -3/56)$ for range for existence and stability of Case 2 solution.

{\bf Case 3}: here we have $\zeta_+$ and all three modes are real and negative everywhere on $\theta > 0$, making entire range of definition for solutions within this class stable
(see Fig.~\ref{fig_3p4_2}(a)).

{\bf Case 4}: similarly to the previous case, all three modes have negative real parts everywhere in the domain of definition (now, for $\theta < -1/4$) making them stable
(see Fig.~\ref{fig_3p4_2}(d)). It is interesting to note that
$\zeta_+$ for {\bf Case 2} and $\zeta_-$ for {\bf Case 4} have exactly the same exponents in the stability analysis -- it seems that sign swaps for $\alpha$ and $\gamma_D$ exactly cancel each other.

To conclude, solutions with $\gamma_D > 0$ exist and stable only for $\alpha < 0$ (Case 2), for $\theta\in(-1/5, -1/8) \cup (-1/16, -3/56)$, which corresponds to
$\xi\in(-27/56, -15/32) \cup (-0.3, 3/8)$ (so that for both positive and negative $\Lambda$). Solutions with $\gamma_D < 0$ exist and stable for both $\alpha > 0$, $\Lambda > 0$ (Case 3) and
$\alpha < 0$, $\Lambda > 0$ ($\xi < -3/2$) (Case 4).

\section{General $(3+D)$-dimensional case}

The procedure for the general case is exactly the same as for previously described $(3+3)$- and $(3+4)$-dimensional cases. So with use of (\ref{3p3_redef1}), the resulting system for
stabilized extra dimensions reads

\begin{equation}
\begin{array}{l}
D(D-1)(D-2)(D-3)\gamma_D^2 + D\zeta\gamma_D (12\theta+1)(D-1) + \zeta^2 (6\theta - \xi) = 0; \\
D(D-1)(D-2)(D-3)(D-4) \gamma_D^2 + \zeta\gamma_D (D-1)(D-2)(24\theta+1) + \zeta^2(24\theta^2 + 12\theta - \xi) = 0.
\end{array} \label{3pD_sys1}
\end{equation}

One can immediately see that due to the multiplier $(D-4)$ in the second equation, we cannot obtain $D=4$ case as a subcase of general one. Following the procedure, we express $\xi$ from one of
the equations and substitute it into another one. The resulting equation is quadratic with respect to $\zeta$ but, unlike previous $(3+3)$- and $(3+4)$-dimensional cases its solutions need radicals to be written explicitly,
 nevertheless, they still could be written in a closed form:

\begin{equation}
\begin{array}{l}
\xi = \dac{12\theta(1+2\theta)\zeta^2 + \gamma_D(D-1)(D-2)(24\theta+1)\zeta + \gamma_D^2(D-1)(D-2)(D-3)(D-4)}{\zeta^2}, \\ \\
\zeta_\pm = \dac{-\gamma_D (D-1) (6D\theta - 24\theta - 1 \pm \sqrt{\mathcal{D}}) }{6\theta (4\theta+1)},~\mbox{where} \\ \\
\mathcal{D} = (D-1) \( (D-1)\gamma_D^2 (6D\theta - 24\theta - 1)^2 + 24\theta(D-2)(D-3)(4\theta+1)   \).
\end{array} \label{3pD_zeta}
\end{equation}

Similarly the the previous cases, we consider separately four cases with different signs for $\alpha$ and $\gamma_D$. But unlike previous cases, now everything depends not only on our parameter $\theta$,
but also on $D$ -- number of extra dimensions. In the previous cases it was easy to plot the resulting curves and find roots/divergences, now it is a bit more complicated.

{\bf Case 1} ($\alpha > 0$, $\gamma_D > 0$) and {\bf Case 3} ($\alpha > 0$, $\gamma_D < 0$) are quite similar: in both we have $\alpha > 0$ so that we can consider only $\theta > 0$ and for solution
to exist we should have $\zeta > 0$. There, we can
see from (\ref{3pD_zeta}) that $\mathcal{D}^2 > (D+1)^2(6D\theta - 24\theta - 1)^2$ which makes $\zeta_+ > 0$ and $\zeta_- < 0$ for both signs of $\gamma_D$. So that for cases 1 and 3 we have a
solution for the entire $\theta > 0$ for $\zeta_+$. When transforming to $\xi$, it corresponds to $\xi > 0$ for Case 1 and $\xi < -D(D-1)/(4(D-2)(D-3))$ for Case 3; one can see that the latter is
different from $(3+4)$-dimensional case while the formed is the same.

Similar analysis could be used to {\bf Case 2} ($\alpha < 0$, $\gamma_D > 0$) and {\bf Case 4} ($\alpha < 0$, $\gamma_D < 0$). Now we have
$\alpha < 0$ so that we consider only $\theta < 0$ and should have $\zeta < 0$ for the solution to exist. Again, comparing discriminant and the free term, we can conclude that Case 4 solutions
exist for $\zeta_-$ and $\theta < -1/4$ while Case 2 solutions exist for all $\theta < 0$ on $\zeta_-$ and $0 > \theta > -1/4$ on $\zeta_+$. One can see that existence  regions for these two cases
 are exactly the same as for $(3+4)$-dimensional case.

\subsection{Linear stability of general $(3+D)$-dimensional solutions}

In the previously described cases for given $\alpha$ and $\gamma_D$ we could
build a solution for any $\theta$ from its region of existence. This general case is different -- $\theta$ is still our parameter but we additionally have number of extra dimensions $D$; also,
solutions for $\zeta_\pm$ (\ref{3pD_zeta}) no more have simple form -- they have radicals which further complicate the analysis. Still, we can formally follow all the steps -- perturb the equations of
motion around the solution, substitute (\ref{3p3_redef1}), calculate eigenvalues of the matrix for perturbation equations and the result will depend on $\theta$ with $D$ and with $\alpha$ and $\gamma_D$ as
parameters. As the resulting eigenvalues being a roots of cubic equation, their implicit analysis is quite troublesome, nevertheless one can demonstrate that they do not have extremum as a function of
$D$ for $D > 4$. With this knowledge at hand we investigate several particular $D$ cases (we used $D = 5,\, 7,\, 10,\, 12$) and, seing no qualitative difference, decide that this is common behavior for
general $D$. This way we demonstrate that in {\bf Case 1} one of the modes is always positive while two others always negative, making solutions from this case unstable -- the same situation as in
previous cases; Fig.~\ref{fig_3p3_2}(a) would serve as a good illustration for this case.
On the other hand, solutions for
{\bf Case 3} are stable, as all three modes are negative for $\theta > 0$ and $\zeta_+$. This is the same as in $(3+4)$-dimensional case, so that Fig.~\ref{fig_3p4_2}(a) is a good illustration.
{\bf Case 4} also proves to be stable for $\zeta_-$ and $\theta < -1/4$ so that Fig.~\ref{fig_3p4_2}(d) is an example.
Finally, {\bf Case 2} is the
most complicated of all four cases - well, just like in $(3+3)$- and $(3+4)$-dimensional cases as well. For $\zeta_-$ branch, one of the modes always has positive real part, making it
unstable -- just like in $(3+4)$-dimensional case (see Fig.~\ref{fig_3p4_2}(b)--(c)).
For $\zeta_+$ branch the situation is as follows: the range of definition is $0 > \theta > -1/4$; out of three modes, one has negative real part for $\theta_1 < \theta < \theta_4$, another mode
has negative real part for $\theta\in(-1/4, \theta_2) \cup (\theta_3, 0)$ and final mode always negative. Here $-1/4 < \theta_1 < \theta_2 < \theta_3 < \theta_4 < 0$ are defined as follows: $\theta_1$
is very lengthy radicand coming from denominator of expression for mode; $\theta_3 = - 1/(4D)$ is another zero of the same denominator; $\theta_2$ and $\theta_4$ are roots of

\begin{equation}
\begin{array}{l}
48D^3 (5D-6) (D+1) (3D^2 - 19D + 32)\theta^4 + 24D^2(31D^4 - 185D^3 + 266D^2 - 272D - 480)\theta^3 + \\ + 12D (21D^4 - 103D^3 + 108D^2 - 268D - 384) \theta^2 + 2(D-2)(2D-3)(9D^2 - 32D + 48)\theta
+ \\ + (D-1)(D+2)(2D-3) = 0.
\end{array} \label{3pD_roots}
\end{equation}

\begin{figure}
\includegraphics[width=0.7\textwidth, angle=0]{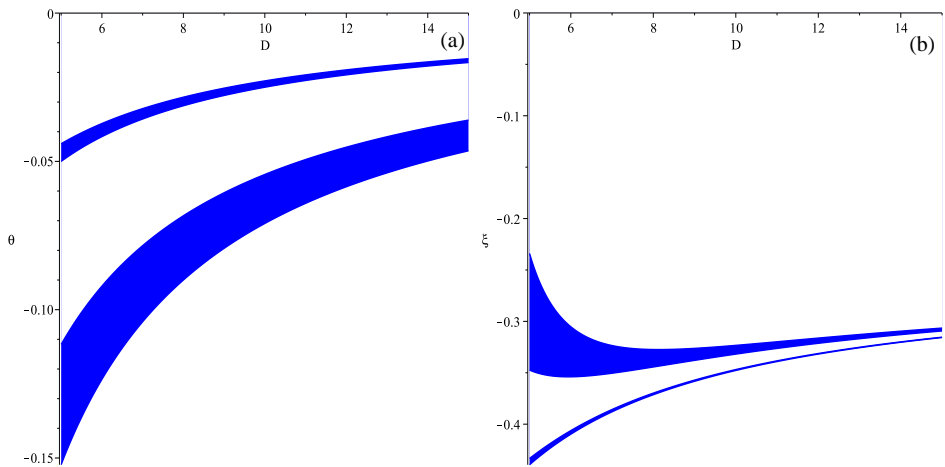}
\caption{Behavior of allowed regions for Case 2 solutions with varying $D$: regions for $\theta$ in (a) panel and for $\xi$ in (b) (see the text for more details).}\label{fig_3pD_1}
\end{figure}

For a solution to be stable, all three modes should be negative;  so the resulting range is intersection of all three; since $-1/4 < \theta_1 < \theta_2 < \theta_3 < \theta_4 < 0$, the resulting
range is $\theta\in(\theta_1, \theta_2)\cup(\theta_3, \theta_4)$ -- see Fig.~\ref{fig_3p4_2}(d) as an example.
Of course, all of these $\theta$'s scales with $D$ and in Fig.~\ref{fig_3pD_1} we presented them for some values of $D$ -- the ranges for $\theta$
are presented in (a) panel; if we convert these ranges into $\xi$, the resulting areas are presented in Fig.~\ref{fig_3pD_1}(b).
One can see from
Fig.~\ref{fig_3pD_1} that the area is decreasing with growth of $D$, making smaller the measure of the parameters when solutions in Case 2 could exist.

\section{Summary}

We realize that the manuscript so far was quite technical and decided to summarize relevant results in a separate section.
Before turning to summarizing the results, there is one more important note we want to make. Through this paper we report existence and stability regions mainly using $\theta$, and it is done for
the reason. Indeed, as we described earlier, $\theta$ is the only ``dynamical'' parameter of the theory (``dynamical'' here means that this parameter vary from one exact solution to another with
$\alpha$, $\gamma_D$ and $D$ remain unchanged). If we have a look on graphs of $\xi$ and $\zeta$ as a functions of $\theta$ we will see that the same value of $\xi$ and $\zeta$ often achieved at
different values of $\theta$: this corresponds to the situation when there are multiple solutions for the same $\alpha$ and $\xi$. For instance, for $(3+3)$-dimensional case there could be up to
three (and for general $(3+D)$ up to four) distinct solutions (with different $H_0$ and $b_0$ each) for a given $\alpha$ and $\xi$. And there will be distinct $\theta$ which corresponds to each of them.
So that if we are interested in finding all possible solutions, we are interested in range of $\theta$; if we are interested in when at least one solution exist, range of $\xi$ would suffice.

With these notes taken, let us move to summarizing the results.

\subsection{Negative curvature of extra dimensions}

In fact, this case was studied quite well in~\cite{CGP1, CGP2} and here it is just one of two subcases of the general $\gamma_D$. Our analysis suggest that the solutions exist in $(3+2)$-
and $(3+3)$-dimensional cases only for $\alpha < 0$ while for $(3+4)$- and higher-dimensional cases -- for both signs of $\alpha$. For $(3+2)$-dimensional case
these solutions are unstable (see Fig.~\ref{fig_3p2_2}(b)) while for higher dimensions the situation becomes more complicated. The solutions with $\alpha < 0$ exist and stable within some ranges
of $\theta$: for $(3+3)$ it is $\theta\in (-1/4, -1/6)$ which corresponds to $\xi < -3/2$ (see Fig.~\ref{fig_3p3_2}(c)); for $(3+4)$ it is $\theta < -1/4$ (and again $\xi < -3/2$); exactly the same
result we obtain for general $(3+D)$ case. Let us note that despite limited range (in $(3+3)$-dimensional case it is even limited from both above and below), it is the entire range of where solution
exist. So that we can conclude that for $\alpha < 0$  all existing solutions are stable.

Now let us turn to $\alpha > 0$ solutions -- they exist only starting from $(3+4)$ and in higher-dimensional cases. Our analysis proves that $(3+4)$- and general $(3+D)$-dimensional cases have exactly the
same description: it is $\zeta_+$ branch which exists (see Fig.~\ref{fig_3p4_1}(f)) and is stable (see Fig.~\ref{fig_3p4_2}(a)) everywhere within range of definition. So that not only for $\alpha < 0$,
but for $\alpha > 0$ all existing solutions are stable as well. So that we can conclude that {\bf all existing solutions with negative curvature and $D>2$ are always stable}.

\subsection{Positive curvature of extra dimensions}

Let us consider separately cases $\alpha > 0$ and $\alpha < 0$, as we did in the negative curvature subsection. The solutions for $\alpha > 0$ always exist (starting from $(3+2)$ and in any higher
dimensions) but are always unstable -- it is what we called Case 1 and one of the nodes always has positive real part in all $D$ cases.
On the contrary, starting from $(3+3)$-dimensional case, $\alpha < 0$
solutions have some range of stability: it is $\theta\in(-\infty, \theta_1) \cup (-1/6, \theta_2) \cup (\theta_5, \theta_4)$, and all $\theta$'s are defined in section dedicated to $(3+3)$-dimensional
model; let us note that the entire $\xi > 0$ is covered within this range of $\theta$. For $(3+4)$-dimensional case the situation is as follows: there are two branches, $\zeta_\pm$, of them $\zeta_-$
defined everywhere at $\theta < 0$ (see Fig.~\ref{fig_3p4_1}(c)) but is stable nowhere (see Fig.~\ref{fig_3p4_2}(b, c)) while $\zeta_+$ is defined at $0 > \theta > -1/4$ and is stable within
$\theta\in(-1/5, -1/8) \cup (-1/16, -3/56)$, which corresponds to $\xi\in(-27/56, -15/32) \cup (-0.3, 3/8)$. Let us note the difference between this and $(3+3)$-dimensional cases -- in the latter
stability range is $\xi > 0$ -- it is unbounded but single-signed. Finally general $(3+D)$-dimensional case follow $(3+4)$-dimensional: for $\alpha < 0$ we have unstable $\zeta_-$ branch and stability of
$\zeta_+$ branch within two intervals; dependence of these intervals on $D$ is depicted in Fig.~\ref{fig_3pD_1}(a) while corresponding range for $\xi$ -- in Fig.~\ref{fig_3pD_1}(b).

\subsection{Concluding remarks}

We summarize all existing and stability ranges in Table~\ref{t1}. Please note that we united results for different branches (when apply), as we are interested in conditions when solutions
exist and are stable, leaving technical details to the main paper. From Table~\ref{t1}
we can clearly see the difference between the cases with positive and negative curvature: solutions with negative curvature are always stable while solutions with positive curvature are stable
only within some range of $\xi$ and the size of this range (``measure of stable trajectories'') strongly decreases with growth of $D$, as indicated in Fig.~\ref{fig_3pD_1}(b). The only exceptions are
$(3+3)$-dimensional case, for which solutions are stable for all $(\alpha < 0, \Lambda < 0)$ and $(3+4)$-dimensional case, where solutions are stable for $\alpha < 0$,
$\xi\in(-27/56, -15/32) \cup (-0.3, 3/8)$.

This can explain why we have not detected this sort of behavior back in~\cite{CGP1, CGP2} -- we studied the behavior numerically, and were interested in large-$D$ asymptotic --
in this case the behavior of models with negative curvature strongly favors detectability (the always stable -- regardless of the parameters) while to find stable solution with positive curvature
we need to fine-tune $\xi$, and with growth of $D$ the size of range for $\xi$ decreases, making it practically impossible to obtain ``correct'' value from this range via random pick.

\begin{table}%[!h]
\caption{Conditions for existing and stability of compactified solutions in different number of extra dimensions $D$}\label{t1}
\centering
\begin{tabular}{|c|c|c|c|c|c|}
\hline
\multicolumn{2}{|c|}{Parameters} & $(3+2)$ & $(3+3)$ & $(3+4)$ & $(3+D)$  \\ \hline
\multirow{2}{*}{$\alpha > 0$, $\gamma_D > 0$} & Exist: & $\xi > 0$ & $\xi > 0$ & $\xi > 0$ & $\xi > 0$ \\ \cline{2-6}
                                              & Stable: & never & never & never & never \\ \hline
\multirow{4}{*}{$\alpha < 0$, $\gamma_D > 0$} & Exist:                   & $\xi > -3/2$           & $\xi > -3/2$                               & $\xi > -3/2$ & $\xi > -3/2$ \\ \cline{2-6}
                                              & \multirow{2}{*}{Stable:} & \multirow{2}{*}{never} & $\xi\in(-0.5448; -0.5)$,  & $\xi\in(-27/56; -15/32)$,  & \multirow{2}{*}{ see Fig.~\ref{fig_3pD_1}(b)} \\
%                                              &                          &                        &                                            &         &  \\
                                              &                          &                        & $\xi > 0$                                  & $\xi\in(-3/10; 3/8)$             &  \\  \hline
\multirow{2}{*}{$\alpha > 0$, $\gamma_D < 0$} & Exist: & no & no & $\xi < 0$ & $\xi < -\dac{D(D-1)}{4(D-2)(D-3)}$ \\ \cline{2-6}
                                              & Stable: & no & no & always & always \\ \hline
\multirow{2}{*}{$\alpha < 0$, $\gamma_D < 0$} & Exist: & $\xi\in(-3/2; -5/6)$ & $\xi < -3/2$ & $\xi < -3/2$ & $\xi < -3/2$ \\ \cline{2-6}
                                              & Stable: & never & always & always & always \\ \hline

\end{tabular}
\end{table}

\begin{figure}
\includegraphics[bb=65 339 459 732, width=0.5\textwidth, angle=0]{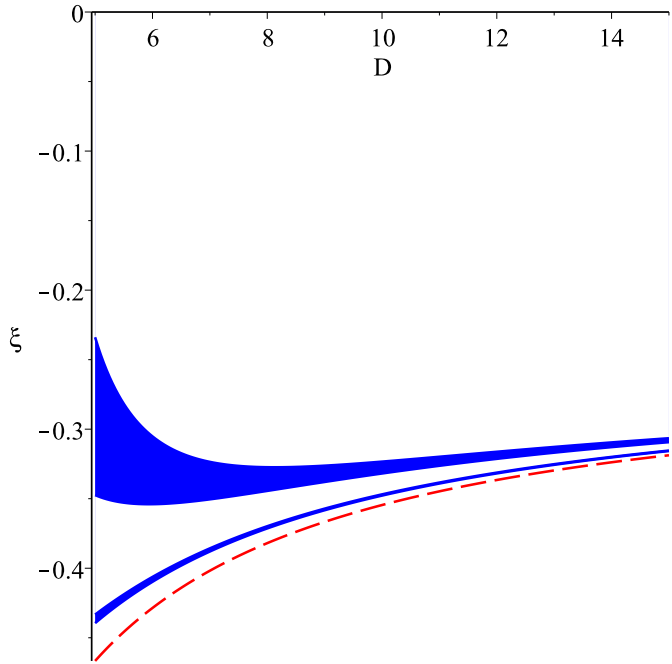}
\caption{Allowed regions for $\xi$ as a functions of the number of extra dimensions (in blue; same as Fig.~\ref{fig_3pD_1}(b)) and separatrix for existence of maximally-symmetric solutions (red dashed line)
 (see the text for more details).}\label{fig8}
\end{figure}

Finally, let us attend one more interesting feature of the obtained solutions. Beforehand, studying the case with negative curvature in~\cite{CGP1, CGP2} we noticed that maximally-symmetric
solutions do not
coexist with compactified ones. In the case of positive curvature they do.
This statement is important so we want to elaborate a bit. If we add criterium for maximally-symmetric solution (\ref{existing_isotrop}) to stability regions on Fig.~\ref{fig_3pD_1}(b), we obtain situation
illustrated in Fig.~\ref{fig8}. In there we took Fig.~\ref{fig_3pD_1}(b) and added exact existence separatrix for maximally-symmetric solution (\ref{existing_isotrop}) as red dashed line.
Since existence criteria for maximally-symmetric solutions is $\xi \geqslant \xi_{iso}$, then one can see that both regions satisfy it.
Unfortunately, the expressions for three out of four $\xi(D)$ curves are
quite cumbersome, so we are left with only one curve -- $\xi_3(D)$ (which is the upper boundary of the bottom (smaller) strip).
It corresponds to $\theta_3 = -1/(4D)$ (see appropriate section) and being substituted to $\xi$ in Eq. (\ref{3pD_zeta}) (alongside
with $\zeta_+$ from the same equation) gives us $\xi_3 = -(D^2+2D-9)/(4D(D-2))$. If we compare it with $\xi_{iso}$ from (\ref{existing_isotrop}) we can see that

 \begin{equation}
\begin{array}{l}
\xi_3 - \xi_{iso} = \dac{3(D-1)}{4D(D+1)(D-2)} > 0,
\end{array} \label{final1}
\end{equation}

\noindent so that $\xi_3 > \xi_{iso}$ always and we can claim that {\bf for $D>2$ there always exist stable compactified solutions with positive curvature which coexist with
maximally-symmetric solutions} -- this
situation is different from the case with negative curvature of extra dimensions. Figure~\ref{fig8} suggests that the statement is even stronger -- that all stable compactified solutions with positive
curvature coexist with maximally-symmetric solutions. To check this we can numerically calculate both $\xi_4$ (lower boundary of the bottom strip) for
 large values of $D$ and
compare it with $\xi_{iso}$ to verify that $\xi_{iso} < \xi_4$. For instance, for $D=100$ we have $\xi_4(D=100) = -0.2599803$ while $\xi_{iso}(D=100) = -0.2600495$, which support our statement
in its stronger form; higher values for $D$ support it as well.

\section{Conclusions}

In this paper we have considered a compactification scenario where stabilisation of extra dimensions occurs due to presence the Gauss-Bonnet term and non-zero spatial curvature.
The sign of spatial curvature can be positive or negative, in the latter case we need additional factorisation to get a compact inner space. Scale factor of the extra dimension space as well as
the effective cosmological constant in three-dimensional ``big'' submanifold are obtained from the system of two algebraic equations. Roots of these equations depend upon coupling constants of the theory, and some combinations of coupling constants can lead either to absence of roots or to negative roots without an adequate physical meaning.

The first goal of the present paper is to describe combinations of coupling constants which allows
a physically acceptable compactification solutions. We show that a product of constants $\xi=\alpha \Lambda$ characterises the existence conditions completely if supplemented by the information about signs of $\alpha$ and the spatial curvature. From our results summarised in Table~\ref{t1} we can see that the existence conditions for both signs of the spatial curvature are not very restrictive. We note also that in the case of negative spatial curvature the compactification solution exist only in those range of $\xi$ where the maximally symmetric solution does not exist. This proves the ``geometrical frustration'' hypothesis for negatively curved inner spaces (see~\cite{CGP1} for details).
We can note that for $\alpha>0$ the compactification solution exists exactly for the range of $\xi$ where maximally symmetric solution does not exist, while for negative $\alpha$ there is a range of $\xi$
where neither compactification nor maximally symmetric solution exist.
However, compactification solutions with positively curved inner spaces can co-exist with the maximally symmetric solution.

The second goal of our paper is to study stability of compactification solutions with respect to homogeneous perturbations of the metric. The first results obtained are the same for both signs of the spatial curvature -- any compactification solution with only $D=2$ extra dimensions is unstable.
This explains why such solutions have never been found in actual numerical integrations of equations of motion. The situation with bigger number of extra dimensions is, however, quite different for negatively and positively curved extra spaces. For the negative curvature extra space,
any solution with $D>2$ extra dimensions is stable. Note also, that for $D>3$ both signs of the constant $\alpha$ are possible for the solution to exist (and, than, to be stable).

For the positive curvature case, on the contrary, stability conditions impose severe restrictions for possible set of the coupling constant of the theory under study. First of all, there are no stable compactification solutions with a positive $\alpha$. Second, for negative $\alpha$ a solution is stable only in rather narrow interval of $\xi$, and the width of this interval decreases with increasing
$D$ (see Fig.~\ref{fig_3pD_1}(b)).

We find that the range of $\xi$ which allows stable compactification solutions is located in zone where maximally symmetric solution also exist. The actual fate of a particular trajectory may depend upon the initial conditions, and this question needs further investigations. In general, positive curvature case seems to be more physically relevant since it leads to compactness of the inner space directly, while addition factorisation is needed for the negative curvature case. Our results
indicate, however, that this advantage is in some sense compensated by the fine-tuning of coupling constants needed for the compactification solution to be stable.  Since this fine-tuning is rather serious for the case of large number of dimensions in the inner space, further Lovelock terms might change these conclusions, and this is the matter of a separate investigation which we plan to provide in future.

\begin{acknowledgments}
The work of A.T. have been supported by the RFBR grant 20-02-00411	and  by the Russian Government Program
of Competitive Growth of Kazan Federal University. 

The work of A.G. was partially supported by FONDECYT project n. 1200293
\end{acknowledgments}


\begin{thebibliography}{99}

\bibitem{LL} D. Lovelock, J. Math. Phys. {\bf 12}, 498 (1971).

\bibitem{Zwie} B. Zwiebach  Phys. Lett. {\bf 156B}, 315 (1985).

\bibitem{BD} D.G. Boulware and S. Deser, Phys. Rev. Lett. {\bf 55}, 2656 (1985).

\bibitem{CGP1} F.~Canfora, A.~Giacomini and S.~A.~Pavluchenko,  Phys.\ Rev.\ D {\bf 88}, 064044 (2013).

\bibitem{CGP2} F.~Canfora, A.~Giacomini and S.~A.~Pavluchenko,  Gen. Rel. Grav. {\bf 46}, 1805 (2014).

\bibitem{CGPT} F.~Canfora, A.~Giacomini, S.~A.~Pavluchenko and A. Toporensky, Gravitation and Cosmology {\bf 24}, 28 (2018).


\bibitem{sch-sch} J. Scherk and J.H. Schwarz, Nucl. Phys. {\bf B81}, 118 (1974).

\bibitem{VSh1} M.A. Virasoro, Phys. Rev. {\bf 177}, 2309 (1969).

\bibitem{VSh2} J.A. Shapiro, Phys. Lett. {\bf 33B}, 361 (1970).

\bibitem{Candelas_etal} P. Candelas, G.T. Horowitz, A. Strominger and E. Witten, Nucl. Phys. {\bf B258}, 46 (1985).

\bibitem{Gross_etal} D.J. Gross, J. Harvey, E. Martinec and R. Rohm, Phys. Rev. Lett. {\bf 54}, 502 (1985).

\bibitem{Lanczos1} C. Lanczos, Z. Phys. {\bf 73}, 147 (1932).

\bibitem{Lanczos2} C. Lanczos, Ann. Math. {\bf 39}, 842 (1938).

\bibitem{zumino} B. Zumino, Phys. Rep. {\bf 137}, 109 (1986).


\bibitem{add_1} F. M${\ddot {\rm u}}$ller-Hoissen, Phys. Lett. {\bf 163B}, 106 (1985).

\bibitem{Deruelle2} N. Deruelle and L. Fari\~na-Busto, Phys. Rev. D {\bf 41}, 3696 (1990).

\bibitem{add_4} F. M${\ddot {\rm u}}$ller-Hoissen, Class. Quant. Grav. {\bf 3}, 665 (1986).


\bibitem{prd09} S.A. Pavluchenko,   Phys. Rev. D {\bf 80}, 107501 (2009).


\bibitem{add_10} J. Demaret, H. Caprasse, A. Moussiaux, P. Tombal, and D. Papadopoulos,  Phys. Rev. D {\bf 41}, 1163 (1990).

\bibitem{add_8} G. A. Mena Marug\'an, Phys. Rev. D {\bf 46}, 4340 (1992).

\bibitem{add13} E. Elizalde, A.N. Makarenko, V.V. Obukhov, K.E. Osetrin, and A.E. Filippov,  Phys. Lett. {\bf B644}, 1 (2007).

\bibitem{MO04} K.I. Maeda and N. Ohta, Phys. Rev. D \textbf{71}, 063520 (2005).

\bibitem{MO14} K.I.~Maeda and N.~Ohta, JHEP {\bf 1406}, 095 (2014).

\bibitem{add_rec_1} J.T. Wheeler, Nucl. Phys. {\bf B268}, 737 (1986).

\bibitem{add_rec_2} R.G. Cai, Phys. Rev. D {\bf 65}, 084014 (2002).

\bibitem{addn_1} T. Torii and H. Maeda, Phys. Rev. D {\bf 71}, 124002 (2005).

\bibitem{addn_2} T. Torii and H. Maeda, Phys. Rev. D {\bf 72}, 064007 (2005).

\bibitem{add_rec_3} D.L. Wilshire. Phys. Lett. {\bf B169}, 36 (1986).

\bibitem{add_rec_4} R.G. Cai, Phys. Lett. {\bf 582}, 237 (2004).

\bibitem{addn_3} J. Grain, A. Barrau, and P. Kanti, Phys. Rev. D {\bf 72}, 104016 (2005).

\bibitem{addn_4} R. Cai and N. Ohta, Phys. Rev. D {\bf 74}, 064001 (2006).

\bibitem{addn_4.1} X.O. Camanho and J.D. Edelstein, Class. Quant. Grav. {\bf 30}, 035009 (2013).


\bibitem{addn_5} H. Maeda, Phys. Rev. D {\bf 73}, 104004 (2006).

\bibitem{addn_6} M. Nozawa and H. Maeda, Class. Quant. Grav. {\bf 23}, 1779 (2006).

\bibitem{addn_7} H. Maeda, Class. Quant. Grav. {\bf 23}, 2155 (2006).

\bibitem{addn_8} M. Dehghani and N. Farhangkhah, Phys. Rev. D {\bf 78}, 064015 (2008).

\bibitem{Deruelle1} N. Deruelle, Nucl. Phys. {\bf B327}, 253 (1989).

\bibitem{mpla09} S.A. Pavluchenko and A.V. Toporensky, Mod. Phys. Lett. {\bf A24}, 513 (2009).

\bibitem{Ivashchuk} V. Ivashchuk, Int. J. Geom. Meth. Mod. Phys. {\bf 07}, 797 (2010) \href{http://arxiv.org/abs/0910.3426v3}{arXiv:0910.3426}.

\bibitem{prd10} S.A. Pavluchenko,   Phys. Rev. D {\bf 82}, 104021 (2010).

\bibitem{grg10}   I.V. Kirnos, A.N. Makarenko, S.A. Pavluchenko, and A.V. Toporensky, General Relativity and Gravitation {\bf 42}, 2633 (2010).

\bibitem{Is86} H. Ishihara, Phys. Lett. \textbf{B179}, 217 (1986).

\bibitem{KPT} I.V. Kirnos, S.A. Pavluchenko, and A.V. Toporensky, Gravitation and Cosmology {\bf 16}, 274 (2010)    \href{http://arxiv.org/abs/1002.4488v2}{arXiv:1002.4488}.

\bibitem{Iv-16} V.D. Ivashchuk, Eur. Phys. J.  C {\bf 76}, 431 (2016); arXiv: 1607.01244v2.

\bibitem{ErIvKob-16} K.K. Ernazarov, V.D. Ivashchuk and A.A. Kobtsev, Grav.  Cosmol., {\bf 22}(3), 245-250 (2016).

\bibitem{CPT1} D. Chirkov, S. Pavluchenko, A. Toporensky, Mod. Phys. Lett. {\bf A29}, 1450093 (2014); \href{http://arxiv.org/abs/1401.2962}{arXiv:1401.2962}.

\bibitem{CST2} D. Chirkov, S. Pavluchenko, A. Toporensky, Gen. Rel. Grav. {\bf 46}, 1799 (2014); \href{http://arxiv.org/abs/1403.4625}{arXiv:1403.4625}.

\bibitem{PT} S.A. Pavluchenko and A.V. Toporensky, Gravitation and Cosmology {\bf 20}, 127 (2014); \href{http://arxiv.org/abs/1212.1386}{arXiv:1212.1386}.

\bibitem{CPT3} D. Chirkov, S. Pavluchenko, A. Toporensky, Gen. Rel. Grav. {\bf 47}, 137 (2015); \href{http://arxiv.org/abs/1501.04360}{arXiv:1501.04360}.

\bibitem{my15} S.A. Pavluchenko,   Phys. Rev. D {\bf 92}, 104017 (2015).

\bibitem{iv16} V. D. Ivashchuk, Eur. Phys. J. C {\bf 76}, 431 (2016).

\bibitem{my16a} S.A. Pavluchenko, Phys. Rev. D {\bf 94}, 024046 (2016).

\bibitem{my18a} S.A. Pavluchenko, Particles {\bf 1}, 36 (2018)  [arXiv:1803.01887].

\bibitem{my18b} S.A. Pavluchenko, Eur. Phys. J. C {\bf 78}, 551 (2018).

\bibitem{my18c} S.A. Pavluchenko, Eur. Phys. J. C {\bf 78}, 611 (2018).

\bibitem{my16b} S.A. Pavluchenko, Phys. Rev. D {\bf 94}, 084019 (2016)

\bibitem{my17a} S.A. Pavluchenko, Eur. Phys. J. C {\bf 77}, 503 (2017).

\bibitem{infl1} S.A. Pavluchenko, Phys. Rev. D {\bf 67}, 103518 (2003).

\bibitem{infl2} S.A. Pavluchenko, Phys. Rev. D {\bf 69}, 021301 (2004).



\bibitem{PT2017} S.A. Pavluchenko and A.V. Toporensky, Eur. Phys. J. C {\bf 78}, 373 (2018).

\bibitem{my18d} S.A. Pavluchenko, Eur. Phys. J. C {\bf 79}, 111 (2019).




\end{thebibliography}
\end{document}